\documentclass[journal]{IEEEtran}
\usepackage{epsbox}
\usepackage[dvips]{graphicx,color}
\usepackage{latexsym}
\usepackage{array}
\begin{document}
\arraycolsep 0.5mm

\newcommand{\bfig}{\begin{figure}[t]}
\newcommand{\efig}{\end{figure}}
\setcounter{page}{1}
\newenvironment{indention}[1]{\par
\addtolength{\leftskip}{#1}\begingroup}{\endgroup\par}
%form: \begin{indention}{2.3cm}
%      \end{indention}
%
%namelist environment
%form: \begin{namelist}{width}
\newcommand{\namelistlabel}[1]{\mbox{#1}\hfill} 
\newenvironment{namelist}[1]{%
\begin{list}{}
{\let\makelabel\namelistlabel
\settowidth{\labelwidth}{#1}
\setlength{\leftmargin}{1.1\labelwidth}}
}{%
\end{list}}
%                                                          %
%\input{ieeecom2e.tex}-------------------------------------%
%                                                          %
\newcommand{\bc}{\begin{center}}  %
\newcommand{\ec}{\end{center}}
\newcommand{\befi}{\begin{figure}[h]}  %
\newcommand{\enfi}{\end{figure}}
\newcommand{\bsb}{\begin{shadebox}\begin{center}}   %
\newcommand{\esb}{\end{center}\end{shadebox}}
\newcommand{\bs}{\begin{screen}}     %
\newcommand{\es}{\end{screen}}
\newcommand{\bib}{\begin{itembox}}   %
\newcommand{\eib}{\end{itembox}}
\newcommand{\bit}{\begin{itemize}}   %
\newcommand{\eit}{\end{itemize}}
\newcommand{\defeq}{\stackrel{\triangle}{=}}
\newcommand{\qed}{\hbox{\rule[-2pt]{3pt}{6pt}}}
\newcommand{\beq}{\begin{equation}}
\newcommand{\eeq}{\end{equation}}
\newcommand{\beqa}{\begin{eqnarray}}
\newcommand{\eeqa}{\end{eqnarray}}
\newcommand{\beqno}{\begin{eqnarray*}}
\newcommand{\eeqno}{\end{eqnarray*}}
\newcommand{\ba}{\begin{array}}
\newcommand{\ea}{\end{array}}
\newcommand{\vc}[1]{\mbox{\boldmath $#1$}}
\newcommand{\lvc}[1]{\mbox{\footnotesize\boldmath $#1$}}
\newcommand{\svc}[1]{\mbox{\scriptsize\boldmath $#1$}}
\newcommand{\wh}{\widehat}
\newcommand{\wt}{\widetilde}
\newcommand{\ts}{\textstyle}
\newcommand{\ds}{\displaystyle}
\newcommand{\scs}{\scriptstyle}
\newcommand{\vep}{\varepsilon}
\newcommand{\rhp}{\rightharpoonup}
\newcommand{\cl}{\circ\!\!\!\!\!-}
\newcommand{\bcs}{\dot{\,}.\dot{\,}}
\newcommand{\eqv}{\Leftrightarrow}
\newcommand{\leqv}{\Longleftrightarrow}
\newcommand{\lan}{\langle}
\newcommand{\ran}{\rangle}
\newcommand{\beqal}{\begin{eqnalp}}
\newcommand{\eeqal}{\end{eqnalp}}
\newcommand{\beqala}{\begin{eqnalp2}}
\newcommand{\eeqala}{\end{eqnalp2}}

\newfont{\bg}{cmr10 scaled \magstep4}
\newcommand{\bigzerol}{\smash{\hbox{\bg 0}}}
\newcommand{\bigzerou}{\smash{\lower1.7ex\hbox{\bg 0}}}
\newcommand{\nbn}{\frac{1}{n}}
\newcommand{\ra}{\rightarrow}
\newcommand{\la}{\leftarrow}
\newcommand{\ldo}{\ldots}
\newcommand{\ep}{\epsilon }
\newcommand{\typi}{A_{\epsilon }^{n}}
\newcommand{\bx}{\hspace*{\fill}$\Box$}
\newcommand{\pa}{\vert}
\newcommand{\ignore}[1]{}
\newtheorem{co}{Corollary} 
\newtheorem{lm}{Lemma} 
\newtheorem{Ex}{Example} 
\newtheorem{Th}{Theorem}
\newtheorem{df}{Definition} 
\newtheorem{pr}{Property} 
\newtheorem{pro}{Proposition} 
\newtheorem{rem}{Remark} 

\newcommand{\lcv}{convex } 

\newcommand{\hugel}{{\arraycolsep 0mm
                    \left\{\ba{l}{\,}\\{\,}\ea\right.\!\!}}
\newcommand{\Hugel}{{\arraycolsep 0mm
                    \left\{\ba{l}{\,}\\{\,}\\{\,}\ea\right.\!\!}}
\newcommand{\HUgel}{{\arraycolsep 0mm
                    \left\{\ba{l}{\,}\\{\,}\\{\,}\vspace{-1mm}
                    \\{\,}\ea\right.\!\!}}
\newcommand{\huger}{{\arraycolsep 0mm
                    \left.\ba{l}{\,}\\{\,}\ea\!\!\right\}}}
\newcommand{\Huger}{{\arraycolsep 0mm
                    \left.\ba{l}{\,}\\{\,}\\{\,}\ea\!\!\right\}}}
\newcommand{\HUger}{{\arraycolsep 0mm
                    \left.\ba{l}{\,}\\{\,}\\{\,}\vspace{-1mm}
                    \\{\,}\ea\!\!\right\}}}

\newcommand{\hugebl}{{\arraycolsep 0mm
                    \left[\ba{l}{\,}\\{\,}\ea\right.\!\!}}
\newcommand{\Hugebl}{{\arraycolsep 0mm
                    \left[\ba{l}{\,}\\{\,}\\{\,}\ea\right.\!\!}}
\newcommand{\HUgebl}{{\arraycolsep 0mm
                    \left[\ba{l}{\,}\\{\,}\\{\,}\vspace{-1mm}
                    \\{\,}\ea\right.\!\!}}
\newcommand{\hugebr}{{\arraycolsep 0mm
                    \left.\ba{l}{\,}\\{\,}\ea\!\!\right]}}
\newcommand{\Hugebr}{{\arraycolsep 0mm
                    \left.\ba{l}{\,}\\{\,}\\{\,}\ea\!\!\right]}}
\newcommand{\HUgebr}{{\arraycolsep 0mm
                    \left.\ba{l}{\,}\\{\,}\\{\,}\vspace{-1mm}
                    \\{\,}\ea\!\!\right]}}

\newcommand{\hugecl}{{\arraycolsep 0mm
                    \left(\ba{l}{\,}\\{\,}\ea\right.\!\!}}
\newcommand{\Hugecl}{{\arraycolsep 0mm
                    \left(\ba{l}{\,}\\{\,}\\{\,}\ea\right.\!\!}}
\newcommand{\hugecr}{{\arraycolsep 0mm
                    \left.\ba{l}{\,}\\{\,}\ea\!\!\right)}}
\newcommand{\Hugecr}{{\arraycolsep 0mm
                    \left.\ba{l}{\,}\\{\,}\\{\,}\ea\!\!\right)}}

\newcommand{\hugepl}{{\arraycolsep 0mm
                    \left|\ba{l}{\,}\\{\,}\ea\right.\!\!}}
\newcommand{\Hugepl}{{\arraycolsep 0mm
                    \left|\ba{l}{\,}\\{\,}\\{\,}\ea\right.\!\!}}
\newcommand{\hugepr}{{\arraycolsep 0mm
                    \left.\ba{l}{\,}\\{\,}\ea\!\!\right|}}
\newcommand{\Hugepr}{{\arraycolsep 0mm
                    \left.\ba{l}{\,}\\{\,}\\{\,}\ea\!\!\right|}}

\newenvironment{jenumerate}
	{\begin{enumerate}\itemsep=-0.25em \parindent=1zw}{\end{enumerate}}
\newenvironment{jdescription}
	{\begin{description}\itemsep=-0.25em \parindent=1zw}{\end{description}}
\newenvironment{jitemize}
	{\begin{itemize}\itemsep=-0.25em \parindent=1zw}{\end{itemize}}
\renewcommand{\labelitemii}{$\cdot$}

\newcommand{\iro}[2]{{\color[named]{#1}#2\normalcolor}}
\newcommand{\irr}[1]{{\color[named]{Red}#1\normalcolor}}
\newcommand{\irg}[1]{{\color[named]{Green}#1\normalcolor}}
\newcommand{\irb}[1]{{\color[named]{Blue}#1\normalcolor}}
\newcommand{\irBl}[1]{{\color[named]{Black}#1\normalcolor}}
\newcommand{\irWh}[1]{{\color[named]{White}#1\normalcolor}}

\newcommand{\irY}[1]{{\color[named]{Yellow}#1\normalcolor}}
\newcommand{\irO}[1]{{\color[named]{Orange}#1\normalcolor}}
\newcommand{\irBr}[1]{{\color[named]{Purple}#1\normalcolor}}
\newcommand{\IrBr}[1]{{\color[named]{Purple}#1\normalcolor}}
\newcommand{\irBw}[1]{{\color[named]{Brown}#1\normalcolor}}
\newcommand{\irPk}[1]{{\color[named]{Magenta}#1\normalcolor}}
\newcommand{\irCb}[1]{{\color[named]{CadetBlue}#1\normalcolor}}
%\newcommand{\irDg}[1]{{\color[named]{DarkSlateGray}#1\normalcolor}}
%
%----------------------------------------------------
%

\newcommand{\Ch}{{\Gamma}}
\newcommand{\Rw}{{W}}

 \newcommand{\Cd}{{\cal R}_{\rm d}(\Ch)}
\newcommand{\Cdi}{{\cal R}_{\rm d}^{\rm (in)}(\Ch)}
\newcommand{\Cdo}{{\cal R}_{\rm d}^{\rm (out)}(\Ch)}

\newcommand{\tCdi}{\tilde{\cal R}_{\rm d}^{\rm (in)}(\Ch)}
\newcommand{\tCdo}{\tilde{\cal R}_{\rm d}^{\rm (out)}(\Ch)}
\newcommand{\hCdo}{  \hat{\cal R}_{\rm d}^{\rm (out)}(\Ch)}

 \newcommand{\Cs}{{\cal R}_{\rm s}(\Ch)}
\newcommand{\Csi}{{\cal R}_{\rm s}^{\rm (in)}(\Ch)}
\newcommand{\Cso}{{\cal R}_{\rm s}^{\rm (out)}(\Ch)}
 \newcommand{\Cds}{{\cal C}_{\rm ds}(\Ch)}
\newcommand{\Cdsi}{{\cal C}_{\rm ds}^{\rm (in)}(\Ch)}
\newcommand{\Cdso}{{\cal C}_{\rm ds}^{\rm (out)}(\Ch)}
\newcommand{\tCdsi}{\tilde{\cal C}_{\rm ds}^{\rm (in)}(\Ch)}
\newcommand{\tCdso}{\tilde{\cal C}_{\rm ds}^{\rm (out)}(\Ch)}
\newcommand{\hCdso}{\hat{\cal C}_{\rm ds}^{\rm (out)}(\Ch)}
 \newcommand{\Css}{{\cal C}_{\rm ss}(\Ch)}
\newcommand{\Cssi}{{\cal C}_{\rm ss}^{\rm (in)}(\Ch)}
\newcommand{\Csso}{{\cal C}_{\rm ss}^{\rm (out)}(\Ch)}
 \newcommand{\Cde}{{\cal R}_{\rm d1e}(\Ch)}
\newcommand{\Cdei}{{\cal R}_{\rm d1e}^{\rm (in)}(\Ch)}
\newcommand{\Cdeo}{{\cal R}_{\rm d1e}^{\rm (out)}(\Ch)}
\newcommand{\tCdei}{\tilde{\cal R}_{\rm d1e}^{\rm (in)}(\Ch)}
\newcommand{\tCdeo}{\tilde{\cal R}_{\rm d1e}^{\rm (out)}(\Ch)}
\newcommand{\hCdeo}{  \hat{\cal R}_{\rm d1e}^{\rm (out)}(\Ch)} 
\newcommand{\Cse}{{\cal R}_{\rm s1e}(\Ch)}
\newcommand{\Csei}{{\cal R}_{\rm s1e}^{\rm (in)}(\Ch)}
\newcommand{\Cseo}{{\cal R}_{\rm s1e}^{\rm (out)}(\Ch)}

\newcommand{\Capa}{C}
\newcommand{\tCapa}{\tilde{C}}
%
%\date{}
\title{
%Separate Coding of Correlated Gaussian Remote Sources
%Distributed Source Coding
Relay Channels with Confidential Messages
%Rate Distortion Region for the Vector Gaussian CEO Problem
}
\author{Yasutada~Oohama,~\IEEEmembership{Member,~IEEE,
}
\thanks{
%Manuscript received xxx, 2001; revised xxx, 2005.
}% 
\thanks{Y. Oohama is with the Department of Information Science 
        and Intelligent Systems,
        University of Tokushima, 
        2-1 Minami Josanjima-Cho, 
        Tokushima 770-8506, Japan.}
}
\markboth{
%IEEE Transactions on Information Theory,~Vol.~XX,No.~Y, 
%~Month~20XX
}
{Oohama: Relay Channels with Confidential Messages
}
% *** Note that you probably will NOT want to include the author's name in ***
% *** the headers of peer review papers.                                   ***

% If you want to put a publisher's ID mark on the page
% (can leave text blank if you just want to see how the
% text height on the first page will be reduced by IEEE)
%\pubid{0000--0000/00\$00.00~\copyright~2002 IEEE}

% \specialpapernotice{(Invited Paper)}

\maketitle

\begin{abstract}
We consider a relay channel where a relay helps the transmission of 
messages from one sender to one receiver. The relay is considered not 
only as a sender that helps the message transmission but as a 
wire-tapper who can obtain some knowledge about the transmitted messages. 
In this paper we study the coding problem of the relay channel under 
the situation that some of transmitted messages are confidential to the 
relay. A security of such confidential messages is measured by the 
conditional entropy. The rate region is defined by the set of 
transmission rates for which messages are reliably transmitted and the 
security of confidential messages is larger than a prescribed level. In 
this paper we give two definition of the rate region.  We first define the 
rate region in the case of deterministic encoder and call it the 
deterministic rate region. Next, we define the rate region in the case 
of stochastic encoder and call it the stochastic rate region. 
We derive explicit inner and outer bounds for the above two rate 
regions and present a class of relay channels where two bounds 
match. Furthermore, we show that stochastic encoder can enlarge the 
rate region. We also evaluate the deterministic rate region of the 
Gaussian relay channel with confidential messages.
%, where two channel outputs are corrupted by 
%additive white Gaussian.  
\end{abstract}


% Will be ignored for other modes.
\IEEEpeerreviewmaketitle

\section{Introduction}

The security of communication systems can be studied from a information 
theoretical viewpoint by regarding them as a kind of cryptosystem in 
which some messages transmitted through communication channel should be 
confidential to anyone except for authorized receivers. The security of 
a communication system was first studied by Shannon \cite{sh} from a 
standpoint of information theory. He discussed a theoretical model of 
cryptosystems using the framework of classical one way noiseless 
channels and derived some conditions for secure communication. 
Subsequently, the security of communication systems based on the 
framework of broadcast channels were studied by Wyner \cite{Wyn1} and 
Csisz{\' a}r and K{\"o}rner \cite{CsiKor1}.  Maurer \cite{Maurer}, 
Ahlswede and Csisz{\' a}r\cite{ac1}, \cite{ac2}, Csisz\'ar and Narayan 
\cite{cn}, and Venkatesan and Anantharam \cite{va} studied the problem 
of public key agreements under the framework of multi-terminal channel 
coding systems. 
%the

Various types of multiterminal channel networks have been investigated 
so far in the field of multi-user information theory. In those networks 
some kind of confidentiality of information transmitted through channels 
is sometimes required from the standpoint of information security. In 
this case it is of importance to analyze the security of communication 
from a viewpoint of multi-user information theory. The author \cite{ohrcc} 
discussed the security of communication using relay 
channels. The author posed and investigate the relay channel with 
confidential messages, where the relay acts as both a helper and a 
wire-tapper. Recently, Liang and Poor \cite{lp} studied the security of 
communication using multiple access channel by formulating and 
investigating the multiple access channel with confidential messages.  

%Most of works on the 
%was  in [4], where the relay node
%acts as both a helper and a wire-tapper.
%
%
%studied the relay channel with confidential messages.
%channel, where a relay helps the transmission of messages from a
%sender to a receiver. 

In this paper we discuss the security of communication for relay channel 
under the framework that the author introduced in \cite{ohrcc}. In the 
relay channel the relay is considered not only as a sender who helps the 
transmission of messages but as a wire-tapper who can learn something 
about the transmitted messages. The coding theorem for the relay channel 
was first established by Cover and El~Gamal \cite{cg}. By carefully 
checking their coding scheme used for the proof of the direct coding 
theorem, we can see that in their coding scheme the relay helps the 
transmission of messages by learning all of them. Hence, this coding 
scheme is not adequate when some messages should be confidential 
to the relay.

The author \cite{ohrcc} studied the security of communication for the 
relay channel under the situation that some of transmitted messages are 
confidential to the relay. For analysis of this situation the author 
posed the communication system called the relay channel with 
confidential messages or briefly said the RCC. In the RCC, a sender 
wishes to transmit two different types of message. One is a message 
called {\it the common message} which is sent to the receiver and the 
relay. The other is a message called {\it the private message} which is 
sent only to the receiver and is confidential to the relay as much as 
possible. The knowledge that the relay gets about private messages is 
measured by the conditional entropy of private messages conditioned by 
channel outputs that the relay observes. The author \cite{ohrcc} defined 
the rate region by the set of transmission rates for which common and 
private messages are transmitted with arbitrary small error 
probabilities and the security of private message measured by the 
conditional entropy per transmission is larger than a prescribed level. 
The author \cite{ohrcc} derived an inner bound of the capacity region 
of the RCC. 

In this paper we study the coding problem of the RCC. In general two 
cases of encoding can be considered in the problem of channel coding. 
One is a case where deterministic encoders are used for transmission of 
messages and the other is a case where stochastic encoders are used. In 
the definition of the rate region by the author \cite{ohrcc}, 
deterministic encoders are implicitly assumed. In this paper we also 
consider the case of stochastic encoders. We define the rate region in 
the case where deterministic encoders are used for transmission and call 
it the deterministic rate region. We further define the rate region in 
the case of stochastic encoders and call it the stochastic rate region. 
We derive explicit inner and outer bounds for the above two rate regions 
and present a class of relay channels where inner and outer bounds match. 
Furthermore, we give another class of relay channels, where the outer 
bound is very close to the inner bound. We also compare the results on 
stochastic and deterministic rate region, demonstrating that stochastic 
encoder can enlarge the rate region. We also study the Gaussian RCC, 
where transmissions are corrupted by additive Gaussian noise. We 
evaluate the deterministic rate region of the Gaussian RCC and derive 
explicit inner and outer bounds. We show that for some class of relay 
channels those two bounds match.  

%As a by-product of the 
%study of RCC, we derive a new upper bound of the capacity of relay 
%channels. We show that for a class of Gaussian relay channels, this 
%upper bound is strictly smaller than the upper bound derived by Cover 
%and El Gamal \cite{cg}. 

Recently, Liang and Veeravalli \cite{lv} and Liang and Krammer \cite{lk} 
posed and investigated a new theoretical model of cooperative 
communication network called the partially/fully cooperative relay 
broadcast channel(RBC). A special case of the partially cooperative RBC 
coincides with the RCC in a framework of communication. However, in the 
problem setup, there seems to be an essential difference between them. 
The formulation of problem in the RBC is focused on an aspect of {\it 
cooperation} in relay channels. On the other hand, the formulation of 
problem by the author \cite{ohrcc} is focused on an aspect of {\it security} 
in relay channels. Cooperation and security are two important features 
in communication networks. It is interesting to note that both 
cooperation and security simultaneously occur in relay communication 
networks. 

%type of communication network which naturally involves those two 
%features. They obtained several capacity results on this communication 
%system. Although their communication framework of partial cooperation 
%and that of the authors  match in the use of relay channel as 
%broadcasting of messages at two destination, our viewpoint and aim are 
%different from theirs.
%
%
%
%
%
%
%
%
%,in which a relay helps the transmission of
%messages from one sender to one receiver. 
%n the other hand the private message is sent only to the receiver and
%should be kept secrete for the relay. 

%For problems not involving secrecy randomized encoding seldom offers any 
%advantage; hence attention is usually restricted to deterministic 
%encoders. Since randomization can increase secrecy, we allow stochastic 
%encoding. 
%For the encoder function we allow a stochastic encoder.  
%specified with 
%A stochastic encoder function 
%with block length $n$ 
%$f_n: {\cal K}_n\times {\cal M}_n$ 
%$\to {\cal X}^n$ 

\section{Relay Channels with Confidential Messages} 

Let ${\cal X},{\cal S},{\cal Y},$ ${\cal Z}$ be finite sets.
The relay channel dealt with in this paper is defined 
by a discrete memoryless channel specified with the following 
stochastic matrix:
\beq
{\Ch} \defeq \{ {\Ch}(y,z\mid x,s)\}_{
(x,s,y,z) 
\in    {\cal X}
\times {\cal S}
\times {\cal Y} 
\times {\cal Z}}\,.
\eeq
Let $X$ be a random variable taking values in ${\cal X}$ and
$X^n=X_{1}X_{2}$ $\cdots X_{n}$ be a random vector taking 
values in ${\cal X}^n$. We write an element of ${\cal X}^n$ as   
${\vc x}=x_{1}x_{2}$ 
$\cdots x_{n}.$
Similar notations are adopted for  
$S,Y,$ and  $Z$.

\begin{figure}[t]
\bc
\includegraphics[width=8.7cm]{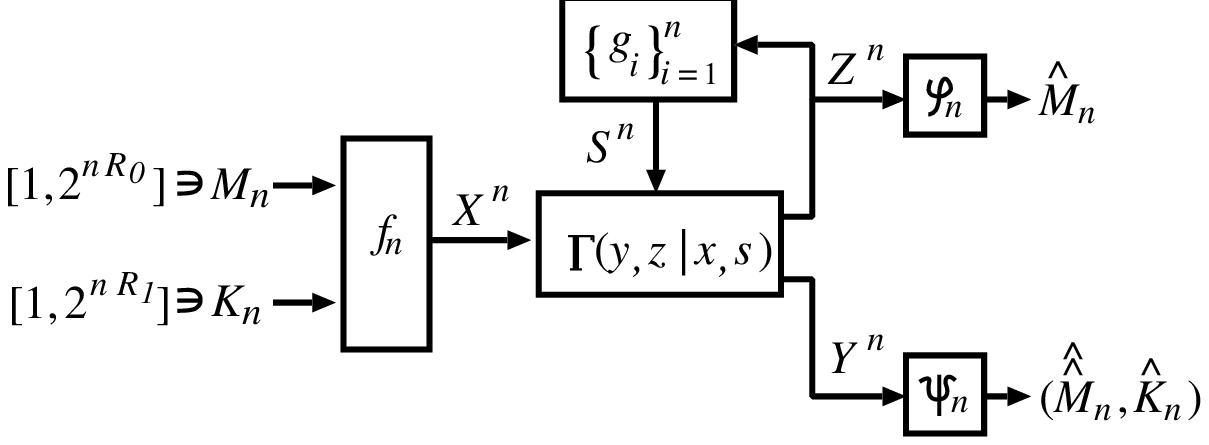}
%\epsfile{file=RccSys2Bw.eps,width=8.7cm}
\caption{Transmission of messages via relay channel 
using $(f_n,\{g_i\}_{i=1}^n,$ $\psi_n,\varphi_n)$.}
\label{fig:blockn} 
\ec
\vspace*{-4mm}
\end{figure}

In the RCC, we consider the following scenario of communication.  
Let $K_n$ and  $M_n$ be uniformly distributed random 
variables taking values in message sets ${\cal K}_n $ and ${\cal M}_n$, 
respectively. The 
random variable $M_n$ is a common message sent to a relay and a receiver. 
The random variable $K_n$ is a private message sent only to the 
receiver and contains an information confidential to the relay. A sender 
transforms $K_n$ and $M_n$ into a transmitted sequence $X^n$ using an 
encoder function $f_n$ and sends it to the relay and the receiver. 
For the encoder function $f_n$, we consider two cases; one is the case 
where $f_n$ is {\it deterministic} and the other is the case 
where $f_n$ is {\it stochastic}. In the former case $f_n$ 
is a one to one mapping 
from ${\cal K}_n\times {\cal M}_n$ to ${\cal X}^n$. 
In the latter case  
$f_n: {\cal K}_n\times {\cal M}_n$ 
$\to {\cal X}^n$ is a stochastic matrix 
defined by 
$$
f_n(k,m)
=\{f_n({\vc x}|k,m)\}_{{\svc x}\in {\cal X}^n },
(k,m)\in{\cal K}_n \times {\cal M}_n\,. 
$$
%\end{document}
Here, $f_n({\vc x}|k,m)$ is the probability that the message $(k,m)$ 
is encoded as a channel input ${\vc x}$.
\begin{figure}[t]
\bc
\includegraphics[width=3.7cm]{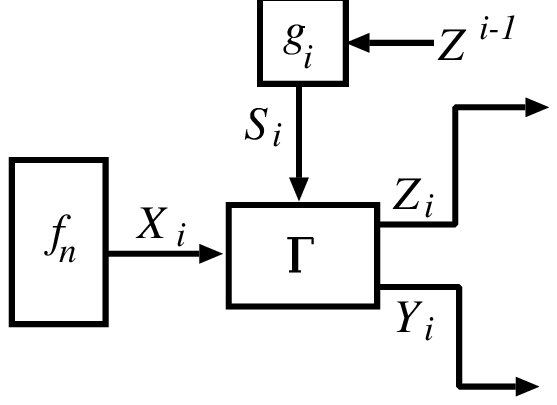}
%\epsfile{file=TransAti3.eps,width=3.7cm}
\caption{Channel inputs and outputs at the $i$th transmission.}
\label{fig:rccblockn} 
\ec
\vspace*{-4mm}
\end{figure}
Channel inputs and outputs at the $i$th transmission is shown 
in Fig. \ref{fig:rccblockn}. At the $i$th transmission, the relay 
observes the random sequence $Z^{i-1}\defeq (Z_1,$ $Z_{2},$ $\cdots, Z_{i-1})$ 
transmitted by the sender through noisy channel, encodes them into 
random variable $S_{i}$ and sends it to the receiver. The relay also 
wishes to decode the common message from observed channel outputs. 
The encoder function at the relay is defined by the sequence of 
functions $\{g_i \}_{i=1}^{n}$.  Each $g_i$ is defined by $g_i: 
{\cal Z}^{i-1}\to {\cal S}$. Note that the channel input $S_{i}$ 
that the relay sends at 
the $i$th transmission depends solely on the output random sequence 
$Z^{i-1}$ that the relay previously obtained as channel outputs. The 
decoding functions at the receiver and the relay are denoted by 
${\psi}_n$ and ${\varphi}_n$, respectively. Those functions are formally 
defined by
$
{\psi}_n: {\cal Y}^{n}   \to {\cal K}_n \times {\cal M}_n\,,
{\varphi}_n: {\cal Z}^{n} \to {\cal M}_n\,.
$
Transmission of messages via relay channel using 
$(f_n,$, $\{g_i \}_{i=1}^{n}$ $\psi_n,\varphi_n)$ is shown in Fig. \ref{fig:blockn}.   
When $f_n$ is a deterministic encoder, error probabilities 
of decoding for transmitted pair $(k,m) \in {\cal K}_n 
\times {\cal M}_n$ are defined by 
\beqa
{\lambda}_{1}^{(n)}(k,m)& \defeq & 
\sum_{\scs ({\svc y},{\svc z}): 
     \atop{\scs \psi_n({\svc y})\neq (k,m)
    }
}\hspace*{-2mm}
\prod_{i=1}^n\Ch\hspace*{-1mm}
\left(y_i,z_i\left|x_i(k,m),g(z^{i-1})\right.\right)\,,
\nonumber\\
{\lambda}_{2}^{(n)}(m)&=& 
\sum_{\scs ({\svc y},{\svc z}):
      \atop
      {\scs \varphi_n({\svc z})\neq m
    }
}%\hspace*{-1mm}
\prod_{i=1}^n\Ch\hspace*{-1mm}
\left(y_i,z_i\left|x_i(k,m),g(z^{i-1})\right.\right)\,,
\nonumber
\eeqa
where $x_i(k,m)$ is the $i$th component of ${\vc x}=f_n(k,m)$.
The average error probabilities 
${\lambda}_{1}^{(n)}$ and ${\lambda}_{2}^{(n)}$ 
of decoding are defined by 
\beqa
{\lambda}_{1}^{(n)} & \defeq &\frac{1}{\pa{\cal K}_n\pa\pa{\cal M}_n\pa}
\sum_{(k,m)\in {\cal K}_n \times {\cal M}_n} 
{\lambda}_{1}^{(n)}(k,m)\,,\\ 
{\lambda}_{2}^{(n)} & \defeq &\frac{1}{\pa {\cal M}_n\pa}\sum_{m\in {\cal M}_n}
{\lambda}_{2}^{(n)}(m)\,,
\eeqa
where $\pa {\cal K}_n \pa$ is a cardinality of the set ${\cal K}_n$.
When $f_n$ is a stochastic encoder, error probabilities 
of decoding for transmitted pair $(k,m) \in {\cal K}_n 
\times {\cal M}_n$ are defined by 
\newcommand{\wid}{\hspace*{-1mm}}
\beqno
\wid& &{\mu}_{1}^{(n)}(k,m)\\
\wid& \defeq &
\sum_{\scs ({\svc x}, {\svc y},{\svc z}): 
     \atop{\scs \psi_n({\svc y})\neq (k,m)
    }
}\hspace*{-2mm}
\prod_{i=1}^n {\Ch}\hspace*{-1mm}
\left(y_i,z_i\left|x_i(k,m),g(z^{i-1})\right.\right)
f_n({\vc x}|k,m)\,,
\\
\wid& &{\mu}_{2}^{(n)}(m)\\
\wid& \defeq & 
\sum_{\scs ({\svc x},{\svc y},{\svc z}): 
      \atop
      {\scs \varphi_n({\svc z})\neq m
    }
}%\hspace*{-1mm}
\prod_{i=1}^n{\Ch}\hspace*{-1mm}
\left(y_i,z_i\left|x_i(k,m),g(z^{i-1})\right.\right)
f_n({\vc x}|k,m)\,.
\eeqno
%where $x_t(k,m)$ is the $t$ th component of $x^n=f(k,m)$.
The average error probabilities ${\mu}_{1}^{(n)}$ 
and ${\mu}_{2}^{(n)}$ of decoding are defined by 
\beqa
{\mu}_{1}^{(n)} & \defeq &\frac{1}{\pa{\cal K}_n\pa\pa{\cal M}_n\pa}
\sum_{(k,m)\in {\cal K}_n \times {\cal M}_n} 
{\mu}_{1}^{(n)}(k,m)\,,\\ 
{\mu}_{2}^{(n)} & \defeq &\frac{1}{\pa{\cal M}_n\pa}\sum_{m\in {\cal M}_n}
{\mu}_{2}^{(n)}(m)\,.
\eeqa
A triple $(R_0,R_1,R_{\rm e})$ is {\it achievable}
if there exists a sequence of quadruples 
$\{(f_n, \{g_i\}_{i=1}^n,$ ${\psi }_n ,{\varphi }_n)\}_{n=1}^{\infty}$ 
such that 
\beqa 
\lim_{n\to\infty}{\lambda}_{1}^{(n)}
&=&\lim_{n\to\infty}{\lambda}_{2}^{(n)}=0\,, 
\nonumber\\
\lim_{n\to\infty} \nbn \log \pa {\cal M}_n \pa & = & R_0 ,
\nonumber
\label{eq:rate1}\\
%	\lim_{n \ra \infty }^{} 
\lim_{n\to\infty} 
\nbn \log \pa {\cal K}_n \pa &= & R_1,
\nonumber
\label{eq:rate2}\\
%	\liminf_{n \ra \infty }^{} 
\lim_{n\to\infty} \nbn H(K_n|Z^{n}) &\geq & R_{\rm e}\,.
\nonumber
\label{eq:rate3}
\eeqa
The set that consists of all achievable rate triple is denoted by 
$\Cd$, which is called the deterministic rate 
region of the RCC. The definition of the stochastic rate region 
$\Cs$ of the RCC is obtained by replacing 
${\lambda}_{1}^{(n)}$ and ${\lambda}_2^{(n)}$ 
in the definition of $\Cd$ by ${\mu}_{1}^{(n)}$ 
and ${\mu}_{2}^{(n)}$, respectively.   

\section{Main Results}

In this section we state our main results. Proofs of the results 
are stated in Section VI. 
\subsection{Deterministic Case}

In this subsection we state our results on inner and outer bounds of 
${\Cd}$. Let $U$ be an auxiliary random variable taking values in finite 
set ${\cal U}$. Define the set of random triples $(U,$ $X,$ $S)$ 
$\in$ ${\cal U}$
$\times{\cal X}$
$\times{\cal S}$
by
\beqno
{\cal P}_1
&\defeq &
\{(U,X,S)%\in {\cal U}\times{\cal X}\times{\cal S}
:
\ba[t]{l} 
\pa {\cal U} \pa 
\leq \pa {\cal X} \pa \pa {\cal S} \pa + 3\,,
\vspace{1mm}\\
\:\:U \ra XS \ra YZ \}\,,
\ea
\eeqno
where    
$U \ra XS \ra YZ$ means that random variables $U,(X,S)$ and 
$(Y,Z)$ form a Markov chain in this order. Set 
\beqno
{\tCdi}
&\defeq &
\ba[t]{l}
\{(R_0,R_1,R_e) : R_0,R_1,R_e \geq 0\,,
\vspace{1mm}\\
\:\:\ba{rcl}
R_0 & \leq &\min \{I(Y;US), I(Z;U|S) \}\,,
\vspace{1mm}\\
R_1 & \leq & I(X;Y|US)\,,
\vspace{1mm}\\
R_{\rm e} & \leq &R_1\,,
\vspace{1mm}\\
R_{\rm e} & \leq & [I(X;Y|US)-I(X;Z|US)]^{+}\,,
\ea
\vspace{1mm}\\
\mbox{ for some }(U,X,S)\in {\cal P}_1\,. \}\,,
\ea
\eeqno
where $[a]^{+}=\max\{0,a\}$. Oohama \cite{ohrcc} 
obtained the following result.
\begin{Th}[Oohama \cite{ohrcc}]{\rm For any relay channel 
$\Ch$, $$\tCdi \subseteq \Cd\,.$$
\label{th:ddirect}
}\end{Th}

%In Oohama \cite{ohrcc} the detail 
%of the proof was omitted. In this paper we outline the proof of 
%Theorem \ref{th:ddirect} by presenting an encoding/decoding scheme 
%that attains $\Cd$. The detail is stated in the appendix.

To state our result on an outer bound of $\Cd$, set
\beqno
\tCdo
&\defeq &
\ba[t]{l}
\{(R_0,R_1,R_{\rm e}) : R_0,R_1,R_{\rm e} \geq 0\,,
\vspace{1mm}\\
\ba[t]{rcl}
R_0 &\leq & \min \{I(Y;US),I(Z;U|S)\}\,,
\vspace{1mm}\\
R_1 &\leq & I(X;YZ|US)\,,
\vspace{1mm}\\
R_0+R_1& \leq &I(XS;Y)\,,
\vspace{1mm}\\
R_{\rm e} &\leq &R_1\,,
\vspace{1mm}\\
R_{\rm e} &\leq & I(X;Y|ZUS)\,,
\ea
\vspace{1mm}\\
\mbox{ for some }(U,X,S)\in {\cal P}_1\,. \}\,.
\ea
\eeqno
Then, we have the following theorem. 
\begin{Th}{\rm For any relay channel $\Ch$,
$$ \Cd \subseteq \tCdo\,.$$
\label{th:dconv}
}\end{Th}

An essential difference between inner and outer bounds 
of $\Cd$ is a gap $\Delta$ given by 
\beqno
\Delta &\defeq&I(X;Y|ZUS)-[I(X;Y|US)-I(X;Z|US)]
\nonumber\\
&=&I(X;ZY|US)-I(X;Y|US)
\nonumber\\
&=&I(X;Z|YUS)\,.
\eeqno
Observe that
\beqa
\Delta
&=&H(Z|YUS)-H(Z|YXUS)
\nonumber\\
&=&H(Z|YUS)-H(Z|YXS)
\label{eqn:delta}\\
&\leq&H(Z|YS)-H(Z|YXS)=I(X;Z|YS)\,,
\nonumber
\eeqa
where (\ref{eqn:delta}) follows from the Markov condition 
$U \to XS$ $ \to YZ$. Hence, $\Delta$ vanishes if the relay 
channel $W=\{{\Ch}(z,y|x,s)$ $\}_{(x,s,y,z)
\in {\cal X}\times{\cal S}\times{\cal Y}\times{\cal Z}}$ 
satisfies the following:
\beq
{\Ch}(z,y|x,s)={\Ch}(z|y,s){\Ch}(y|x,s).
\eeq
The above condition is equivalent to the condition that
$X,S,Y,Z$ form a Markov chain $X \to SY\to Z$ in 
this order. Cover and El. Gamal \cite{cg} called this 
relay channel the reversely degraded relay channel. 
On the other hand, we have   
\beqa
& &I(X;Y|ZUS)
\nonumber\\
&=&H(Y|ZUS)-H(Y|ZXUS)
\nonumber\\
&\leq &H(Y|ZS)-H(Y|ZXS)=I(X;Y|ZS)\,,
\label{eqn:delta2}
\eeqa
where (\ref{eqn:delta2}) follows from the 
Markov condition $U \to XSZ\to Y$. 
The quantity $I(X;Y|ZUS)$ vanishes if the relay 
channel $\Ch$ satisfies the following:
\beq
{\Ch}(z,y|x,s)={\Ch}(y|z,s){\Ch}(z|x,s).
\label{eqn:degraded}
\eeq
Hence, if the relay channel $\Ch$ satisfies (\ref{eqn:degraded}), then 
$R_{\rm e}$ should be zero. This implies that no security on the private 
messages is guaranteed. 
%for the degraded relay channel. 
The condition 
(\ref{eqn:degraded}) is equivalent to the condition that $X,S,Y,Z$ form 
a Markov chain $X \to SZ\to Y$ in this order. Cover and El. Gamal 
\cite{cg} called this relay channel the degraded relay channel. 
Summarizing the above arguments, we obtain the following two corollaries. 

\begin{co}{\rm For the reversely degraded relay channel $\Ch$, 
we have 
$$
\tCdi=\Cd=\tCdo\,.
$$
}
\end{co}

\begin{co}{\rm In the deterministic case, 
if the relay channel $\Ch$ is degraded, then no security 
on the private messages is guaranteed.
}
\end{co}

Next, we derive another inner bound 
and two other outer bounds of $\Cd$. 
Define a set of 
random triples $(U,$$X,$ $S)$ 
$\in$ ${\cal U}$
$\times{\cal X}$
$\times{\cal S}$
by
\beqno
{\cal P}_2
&\defeq &
\{(U,X,S)%\in {\cal U}\times{\cal V}\times{\cal X}\times{\cal S}
: 
\ba[t]{l} 
\pa {\cal U} \pa 
\leq \pa {\cal Z} \pa \pa {\cal X} \pa \pa {\cal S} \pa + 3\,,
\vspace{1mm}\\
U \ra XSZ \ra Y \}\,.
\ea
\eeqno
It is obvious that ${\cal P}_1\subseteq{\cal P}_2$. For given 
$(U,X,S)$ 
$\in$ ${\cal U}$
$\times{\cal X}$
$\times{\cal S}$,
set
\beqno
& & {\cal R}(U,X,S|\Ch)
\\
&\defeq &
\ba[t]{l}
\{(R_0,R_1,R_{\rm e}): R_0,R_1,R_{\rm e}\geq 0\,,
\vspace{1mm}\\
\ba[t]{rcl}
R_0&\leq &\min \{I(Y;US), I(Z;U|S)\}\,,
\vspace{1mm}\\
R_0+ R_1 &\leq &I(X;Y|US)
\vspace{1mm}\\
& &+\min \{I(Z;U|S),I(Y;US) \}\,,
\vspace{1mm}\\
R_{\rm e} &\leq & R_1\,, 
\vspace{1mm}\\
R_{\rm e} &\leq & [I(X;Y|US)-I(X;Z|US)]^{+}\,.\}\,.
\ea
\ea
\eeqno
Furthermore, set 
\beqno
\Cdi& \defeq &
\bigcup_{(U,X,S)\in {\cal P}_1}
{\cal R}(U,X,S|\Ch)\,,
\\
\Cdo& \defeq &
\bigcup_{(U,X,S)\in {\cal P}_2}
{\cal R}(U,X,S|\Ch)\,.
\eeqno
Then, we have the following.
\begin{Th}\label{th:dregion} {\rm For any relay channel $\Ch$, 
$$\Cdi\subseteq \Cd \subseteq \Cdo\,.$$
}
\end{Th}

Now we consider the case where 
the relay channel $\Ch$ satisfies 
\beq
{\Ch}(y,z|x,s)={\Ch}(y|x,s){\Ch}(z|x).
\label{eqn:indep}
\eeq
%then $\hCdo =\Cdo$.
The above condition on $\Ch$ is equivalent to the 
condition that $X,S,Y,Z$ satisfy the following 
two Markov chains:
\beqno
Y \to XS \to Z\,, S\to X \to Z\,.
\eeqno
The first condition is equivalent to that 
$Y$ and $Z$ are conditionally independent given $SX$ and 
the second is equivalent to that $Z$ and $S$ are 
conditionally independent given $X$. 
%this condition the CI condition. 
%The CI condition is equivalent to the condition that 
%$X,S,Y,Z$ form a Markov chain $Z \to SX\to Y$ in this order. 
%In this for this class of relay channels.

We say that the relay channel $\Ch$ belongs to the independent 
class if it satisfies (\ref{eqn:indep}). 
%We 
For the independent class of relay channels, we 
derive an outer 
bound of $\Cd$. 
To state our result, set 
\beqno 
& &\hCdo
\\
&\defeq &
\ba[t]{l}
\{(R_0,R_1,R_{\rm e}): R_0,R_1,R_{\rm e}\geq 0\,,
\vspace{1mm}\\
\ba[t]{rcl}
 R_0&\leq &\min \{I(Y;US),I(Z;U|S)\}\,,
\vspace{1mm}\\
 R_0+ R_1 &\leq &I(X;Y|US)+\left[\zeta(U,S,Y,Z)\right]^{+}
\vspace{1mm}\\
& &+\min \{I(Z;U|S),I(Y;US)\}\,,
\vspace{1mm}\\
R_{\rm e} &\leq & R_1\,, 
\vspace{1mm}\\
R_{\rm e}&\leq & 
[I(X;Y|US)-I(X;Z|US)
\vspace{1mm}\\
&  &+\zeta(U,S,Y,Z)]^{+}\,,
\vspace{1mm}\\
& &\mbox{ for some }(U,X,S) \in {\cal P}_1\,. \}\,,
\ea
\ea
\eeqno
where we set 
\beqno
\zeta(U,S,Y,Z) 
&\defeq&   
I(XS;Y|U)-I(XS;Z|U)
\nonumber\\
& &-[I(X;Y|US)-I(X;Z|US)]
\nonumber\\
&=&I(S;Y|U)-I(S;Z|U)
\nonumber\\
&=&H(S|ZU)-H(S|YU)\,.
\eeqno
The quantity $\zeta(U,S,Y,Z)$ 
satisfies the following. 
\begin{pr} For any $(U,X,S)\in {\cal P}_2$,   
\beq
\zeta(U,S,Y,Z) \leq I(XS;Y|Z)\,.
\label{eqn:zeta}
\eeq
\end{pr}
{\it Proof:} We have the following chain of inequalities:
\beqa
& &\zeta(U,S,Y,Z)
\nonumber\\
&= & H(S|ZU)-H(S|YU)
\nonumber\\
&\leq& H(S|ZU)-H(S|YZU)
\nonumber\\
& = & I(S;Y|ZU)
\nonumber\\
& =   & H(Y|ZU)-H(Y|ZUS)
\nonumber\\
&\leq & H(Y|Z)-H(Y|ZUS)
\nonumber\\
&\leq & H(Y|Z)-H(Y|ZXSU)
\nonumber\\
&= & H(Y|Z)-H(Y|ZXS)=I(XS;Y|Z),
%\nonumber
\eeqa
where the last equality follows from 
the Markov condition $U\to$ 
$ZXS \to Y.$
\hfill\QED  

Our result is the following. 
\begin{Th}\label{th:thDI}{\rm If $\Ch$ belongs to the independent 
class, we have 
$$
\Cd \subseteq \hCdo\,.
$$
}
\end{Th}
%We call 

\subsection{Stochastic Case}

%Next, we state results 
%on the stochastic capacity region. 
%Let ${\cal Q}_1$ be the set 
%of all the random triples $(U,$ $V,$ $X,$ $S)$ 
%that satisfy the following.    
In this subsection we state our results on inner and outer bounds 
of ${\Cs}$. Define two sets of random quadruples $(U,$$V,$$X,$$S)$ 
$\in$ ${\cal U}$
$\times{\cal V}$
$\times{\cal X}$
$\times{\cal S}$
by
\beqno
{\cal Q}_1
&\defeq &
\{(U,V,X,S)%\in {\cal U}\times{\cal V}\times{\cal X}\times{\cal S}
:
\ba[t]{l} 
\pa {\cal U} \pa 
\leq  \pa {\cal X} \pa \pa {\cal S} \pa + 3\,,
\vspace{1mm}\\
\pa {\cal V} \pa 
  \leq \left(\pa {\cal X} \pa \pa {\cal S}\pa\right)^2  
  +4\pa {\cal X} \pa \pa {\cal S} \pa + 3\,,
\vspace{1mm}\\
U \ra V \ra XS \ra YZ\,,
\vspace{1mm}\\
US \ra V \ra X 
\}\,,
\ea
\\
{\cal Q}_2
&\defeq &
\{(U,V,X,S)
%\in {\cal U}\times{\cal V}\times{\cal X}\times{\cal S}
:
\ba[t]{l} 
\pa {\cal U} \pa 
\leq \pa {\cal Z} \pa\pa {\cal X} \pa \pa {\cal S} \pa + 3\,,
\vspace{1mm}\\
\pa {\cal V} \pa 
  \leq \left(\pa {\cal Z} \pa\pa {\cal X} \pa \pa {\cal S}\pa\right)^2  
  +4\pa {\cal Z} \pa\pa {\cal X} \pa \pa {\cal S} \pa + 3\,,
\vspace{1mm}\\
U \ra V \ra XSZ \ra Y\,,
\vspace{1mm}\\
US \ra VX\ra Z\,,
\vspace{1mm}\\
US \ra V \ra X \}\,.
\ea
\eeqno
It is obvious that ${\cal Q}_1\subseteq{\cal Q}_2$. For given 
$(U,V,X,S)$ 
$\in$ ${\cal U}$
$\times{\cal V}$
$\times{\cal X}$
$\times{\cal S}$,
set
\beqno
& & {\cal R}(U,V,X,S|\Ch)
\\
&\defeq &
\ba[t]{l}
\{(R_0, R_1, R_{\rm e}): R_0, R_1, R_{\rm e}\geq 0 \,, 
\vspace{1mm}\\
\ba[t]{rcl}
R_0&\leq &\min \{I(Y;US),I(Z;U|S)\}\,,
\vspace{1mm}\\
R_0+ R_1 &\leq &I(V;Y|US)
\vspace{1mm}\\
& &+\min \{I(Y;US),I(Z;U|S)\}\,,
\vspace{1mm}\\
R_{\rm e} &\leq & R_1\,,
\vspace{1mm}\\
R_{\rm e}&\leq & [I(V;Y|US)-I(V;Z|US)]^{+}\,.\}\,.
\ea
\ea
\eeqno
Set 
\beqno
\Csi&\defeq&
\bigcup_{(U,V,X,S)\in {\cal Q}_1}
{\cal R}(U,V,X,S|\Ch)\,,
\\
\Cso&\defeq&
\bigcup_{(U,V,X,S)\in {\cal Q}_2}
{\cal R}(U,V,X,S|\Ch)\,.
\eeqno
Then, we have the following.
\begin{Th}\label{th:sregion} {\rm For any relay channel $\Ch$, 
$$\Csi\subseteq \Cs \subseteq \Cso \,.$$
}
\end{Th}

Similarly to the deterministic case, we estimate the quantity 
$I(X;Y|US)-I(X;Z|US)$. %.in the case where $\Ch$ is degraded. 
We have the following chain of inequalities:
\beqa
&     & I(V;Y|US)-I(V;Z|US)
\nonumber\\
&\leq & I(V;YZ|US)-I(V;Z|US)
\nonumber\\
&= & I(V;Y|ZUS)
\nonumber\\
& = & H(Y|ZUS)-H(Y|ZVUS)
\nonumber\\
&\leq & H(Y|ZS)-H(Y|ZXVUS)
\nonumber\\
&= &H(Y|ZS)-H(Y|ZXS)=I(X;Y|ZS)\,,
\label{eqn:delta3}
\eeqa
where (\ref{eqn:delta3}) follows from 
the Markov condition 
$$
U \to V \to XSZ\to Y\,.
$$ 
Then, if $\Ch$ is degraded, for any 
$(U,V,S,X)\in {\cal Q}_2$, we have 
$$
I(V;Y|US)-I(V;Z|US)\leq 0\,. 
$$
Hence, if the relay channel $\Ch$ 
%satisfies (\ref{eqn:degraded}), 
is degraded, 
then $R_{\rm e}$ should be zero. This implies that no security on 
the private messages is guaranteed for the degraded relay channel.
Thus, we obtain the following corollary.

\begin{co}{\rm 
When the relay channel $\Ch$ is degraded, 
no security on the private messages is guaranteed 
even if $f_n$ is a stochastic encoder.    
}
\end{co}

%According to Cover and El Gamal \cite{cg}, the capacity of the relay 
%channels is not found except for some special classes of 
%relay channels called physically degraded relay channel, reversely
%degraded relay channel, relay channel with feedback and 
%deterministic relay channel. It also seems very difficult 
%to find an explicit capacity region 
%of the general RCC.

\section{Secrecy Capacities of the RCC} 

In this section we derive an explicit inner and outer bounds of the secrecy
capacity region by using the results in the previous section. 

\subsection{Deterministic Case}

We first consider the case where $f_n$ is a deterministic encoder. 
The secrecy capacity region ${\Cds}$ for the RCC is defined by 
\beq
\Cds = \{(R_0,R_1):(R_0,R_1,R_1)\in \Cd \}\,.
\label{df:Cs_region}
\eeq

From Theorems \ref{th:ddirect} and \ref{th:dconv}, we obtain the 
following corollary. 
\begin{co} For any relay channel $\Ch$, 
$$
\Cdsi \subseteq \Cds \subseteq \tCdso\,, 
$$
where 
\beqno
\Cdsi
&\defeq &
\ba[t]{l}
\{(R_0,R_1) : R_0,R_1\geq 0\,,
\vspace{1mm}\\
\:\:R_0 \leq \min \{I(Y;US), I(Z;U|S)\}\,,
\vspace{1mm}\\
\:\:R_1 \leq [I(X;Y|US)-I(X;Z|US)]^{+}\,,
\vspace{1mm}\\
\mbox{ for some }(U,X,S)\in {\cal P}_1\,. \}\,,
\ea
\\
\tCdso
&\defeq &
\ba[t]{l}
\{(R_0,R_1) : R_0,R_1\geq 0\,,
\vspace{1mm}\\
\:\:R_0 \leq \min \{I(Y;US), I(Z;U|S)\}\,,
\vspace{1mm}\\
\:\:R_1 \leq I(X;Y|ZUS)\,,
\vspace{1mm}\\
\mbox{ for some }(U,X,S)\in {\cal P}_1\,. \}\,.
\ea
\eeqno
In particular, if $\Ch$ is reversely degraded, we have 
$$
\Cdsi=\Cds = \tCdso. 
$$
\end{co}

From Theorem \ref{th:dregion}, we obtain the following corollary. 
\begin{co} For any relay channel $\Ch$, 
$$
\Cdsi \subseteq \Cds \subseteq \Cdso\,, 
$$
where
\beqno
\Cdso
&\defeq&
\ba[t]{l}
\{(R_0,R_1) : R_0,R_1\geq 0\,,
\vspace{1mm}\\
\:\:R_0 \leq \min \{I(Y;US), I(Z;U|S)\}\,,
\vspace{1mm}\\
\:\:R_1 \leq [I(X;Y|US)-I(X;Z|US)]^{+}\,,
\vspace{1mm}\\
\mbox{ for some }(U,X,S)\in {\cal P}_2\,.\}\,.
\ea
\eeqno
%In particular, if $\Ch$ is degraded, 
%we have 
%$$
%\Cdsi=\Cds=\Cdso. 
%$$
\end{co}

From Theorem \ref{th:thDI}, we obtain the following corollary. 
\begin{co} If $\Ch$ belongs to the independent class, we have
$$
\Cds \subseteq \hCdso\,, 
$$
where
\beqno
\hCdso
&\defeq&
\ba[t]{l}
\{(R_0,R_1) : R_0,R_1\geq 0\,, 
\vspace{1mm}\\
\:\:R_0 \leq \min \{I(Y;US), I(Z;U|S)\}\,,
\vspace{1mm}\\
\:\:R_1 \leq [I(X;Y|US)-I(X;Z|US)
\vspace{1mm}\\
\:\:\qquad\qquad + \zeta(U,S,Y,Z)]^{+}\,,
\vspace{1mm}\\
\mbox{ for some }(U,X,S)\in {\cal P}_1\,.\}\,.
\ea
\eeqno
\end{co}

\begin{figure}[t]
\bc
\includegraphics[width=8.0cm]{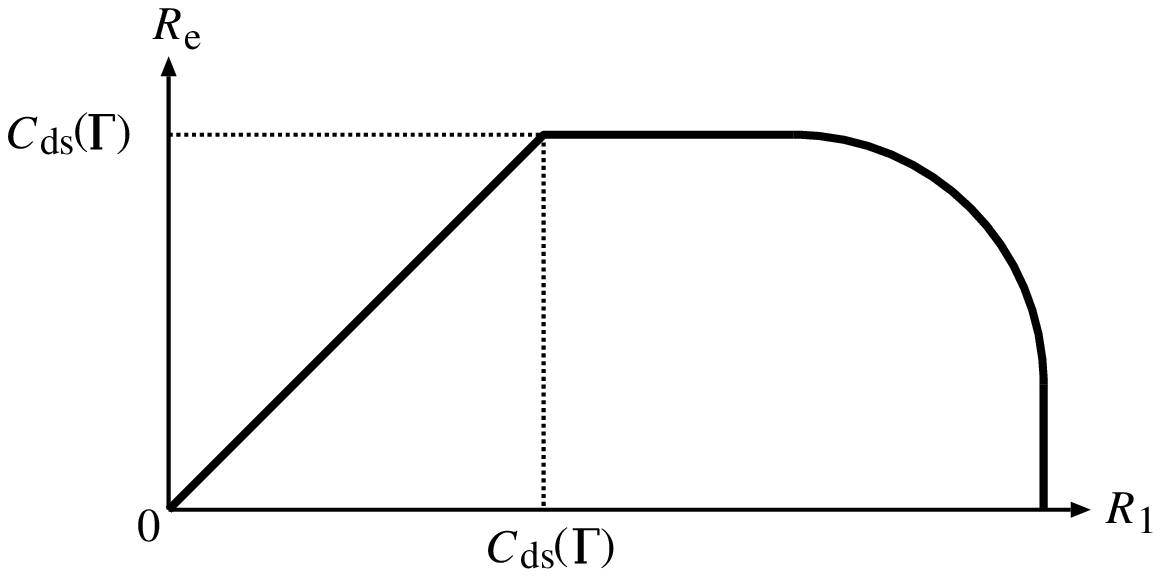}
\caption{ The region $\Cde$ 
and the secrecy capacity $C_{\rm ds}(\Ch)$.
} 
\label{fig:Cs_region}
\ec
%\vspace*{-5mm}
\end{figure}

Now, we consider the special case of no common message. 
Set 
\beqno
{\Cde}\defeq \{(R_1,R_{\rm e}):(0,R_1,R_{\rm e})\in \Cd \}
\eeqno
and define the secrecy capacity by 
\beqno
C_{\rm ds}(\Ch)\defeq 
 \max_{(R_1,R_1)\in \Cde}R_1
=\max_{(0,R_1)\in \Cds}R_1\,.
\eeqno
Typical shape of the region $\Cde$ and the secrecy capacity 
$C_{\rm ds}(\Ch)$ is shown in Fig. \ref{fig:Cs_region}.

From Theorems \ref{th:ddirect} and \ref{th:dconv}, 
we have the following corollary. 
\begin{co} For any relay channel $\Ch$, 
$$
\tCdei \subseteq \Cde \subseteq \tCdeo\,, 
$$
where 
\beqno
\tCdei
&\defeq &
\ba[t]{l}
\{(R_1,R_{\rm e}): R_1,R_{\rm e} \geq 0\,, 
\vspace{1mm}\\
\ba[t]{rcl}
\:\:R_1 &\leq &I(X;Y|US)\,,
\vspace{1mm}\\
\:\:R_{\rm e} &\leq & R_1\,,
\vspace{1mm}\\
\:\:R_{\rm e} &\leq & [I(X;Y|US)-I(X;Z|US)]^{+}\,,
\vspace{1mm}\\
& &\mbox{ for some }(U,X,S)\in {\cal P}_1\,. \}\,,
\ea
\ea
\\
\tCdeo
&\defeq &
\ba[t]{l}
\{(R_1,R_{\rm e}): R_1,R_{\rm e} \geq 0\,, 
\vspace{1mm}\\
\ba[t]{rcl}
\:\:R_1 &\leq &I(X;YZ|US)\,,
\vspace{1mm}\\
\:\:R_{\rm e} &\leq & R_1\,,
\vspace{1mm}\\
\:\:R_{\rm e} &\leq & I(X;Y|ZUS)\,,
\vspace{1mm}\\
& &\mbox{ for some }(U,X,S)\in {\cal P}_1\,. \}\,.
\ea
\ea
\eeqno
Furthermore, 
\beqno
&     & \max_{(U,X,S)\in {\cal P}_1}
\left[I(X;Y|US)-I(X;Z|US)\right]^{+}
\nonumber\\
&\leq & C_{\rm ds}(\Ch)
\nonumber\\
&\leq & \max_{(U,X,S)\in {\cal P}_1}I(X;Y|ZUS)\,.
\eeqno
In particular, if $\Ch$ is reversely degraded, we have 
$$
\tCdei=\Cde = \tCdeo 
$$
and 
\beqno
C_{\rm ds}(\Ch)
&= & \max_{(U,X,S)\in {\cal P}_1}
\left[I(X;Y|US)-I(X;Z|US)\right]\,.
\eeqno
\end{co}

Next, we state a result which is obtained as a corollary of 
Theorem \ref{th:dregion}. Set %To describe our result, set
\beqno
& &{\cal R}_{\rm 1e}(U,X,S|\Ch)
\\
&\defeq & {\cal R}(U,X,S|\Ch)\cap \{(R_0,R_1,R_{\rm e}): R_0=0\}
\\
&=&
\ba[t]{l}
\{(R_1,R_{\rm e}): R_1, R_{\rm e}\geq 0\,,
\vspace{1mm}\\
\ba[t]{rcl}
\:\:R_1 &\leq &I(X;Y|US)
\vspace{1mm}\\
              &     &+ \min \{I(Y;US),I(Z;U|S)\}\,,
\vspace{1mm}\\
\:\:R_{\rm e} &\leq & R_1\,,
\vspace{1mm}\\
\:\:R_{\rm e} &\leq & [I(X;Y|US)-I(X;Z|US)]^{+}\,.\}
\ea
\ea
\eeqno
and 
\beqno
\Cdei&\defeq&
\bigcup_{(U,X,S)\in {\cal P}_1}
{\cal R}_{\rm 1e}(U,X,S|\Ch)\,,
\\
\Cseo&\defeq&
\bigcup_{(U,X,S)\in {\cal P}_2}
{\cal R}_{\rm 1e}(U,X,S|\Ch)\,.
\eeqno
Then, we have the following. %corollary.
\begin{co} For any relay channel $\Ch$, 
$$
\Cdei \subseteq \Cde \subseteq \Cdeo\,. 
$$
Furthermore, 
\beqno
C_{\rm ds}(\Ch)
&\leq & \max_{(U,X,S)\in {\cal P}_2}
\left[I(X;Y|US)-I(X;Z|US)\right]^{+}\,.
\eeqno
\end{co}

Finally we state a result which is obtained as a corollary of 
Theorem \ref{th:thDI}. Set %To describe our result, set
\beqno
%& &
\hCdeo
%\\
&=&
\ba[t]{l}
\{(R_1,R_{\rm e}): R_1, R_{\rm e}\geq 0\,,
\vspace{1mm}\\
\ba[t]{rcl}
\:\:R_1 &\leq &I(X;Y|US)+\left[\zeta(U,S,Y,Z)\right]^{+}
\vspace{1mm}\\
              &     &+ \min \{I(Y;US),I(Z;U|S)\}\,,
\vspace{1mm}\\
\:\:R_{\rm e} &\leq & R_1\,,
\vspace{1mm}\\
\:\:R_{\rm e} &\leq & [I(X;Y|US)-I(X;Z|US)
\vspace{1mm}\\
& & +\zeta(U,Y,S,Z)]^{+}\,,
\vspace{1mm}\\
& & \mbox{ for some }(U,X,S)\in{\cal P}_1\,.\}
\ea
\ea
\eeqno
Then we have the following. %corollary.
\begin{co}{\rm If $\Ch$ belongs to the independent class, we have  
$$
\Cde \subseteq \hCdeo\,.
$$
}
Furthermore, 
\beqno
C_{\rm ds}(\Ch)
&\leq & \max_{(U,X,S)\in {\cal P}_1}
      \left[I(X;Y|US)-I(X;Z|US)\right.
\\
& &\qquad\qquad \quad\left. + \zeta(U,S,Y,Z) \right]^{+}
\\
&=&\max_{(U,X,S)\in {\cal P}_1}
   \left[I(XS;Y|U)-I(XS;Z|U)\right]^{+}\,.
\eeqno

\end{co}

\subsection{Stochastic Case}

The stochastic secrecy capacity region ${\Css}$ for 
the RCC is defined by 
\beq
\Css = \{(R_0,R_1):(R_0,R_1,R_1) 
\in \Cs \}\,.
\label{df:Css_region}
\eeq
To describe our result set  
\beqno
& &{\cal C}_{\rm s}(U,V,X,S|\Ch)
\\
&\defeq& {\cal R}(U,V,X,S|\Ch)\cap
\{ (R_0,R_1,R_{\rm e}): R_1=R_{\rm e} \}
\\
&=&
\ba[t]{l}
\{(R_0,R_1) : R_0,R_1\geq 0\,,
\vspace{1mm}\\
\:\:R_0 \leq \min \{I(Y;US),I(Z;U|S)\}\,,
\vspace{1mm}\\
\:\:R_1 \leq [I(V;Y|US)-I(V;Z|US)]^{+}\,\}
\vspace{1mm}\\
\ea
\eeqno
and 
\beqno
\Cssi&\defeq &
\bigcup_{(U,V,X,S)\in {\cal Q}_1} {\cal C}_{\rm s}(U,V,X,S|\Ch)\,,
\\ 
\Csso&\defeq &
\bigcup_{(U,V,X,S)\in {\cal Q}_2} {\cal C}_{\rm s}(U,V,X,S|\Ch)\,.
\eeqno
From Theorem \ref{th:sregion}, we obtain the following corollary. 
\begin{co} For any relay channel $\Ch$, 
$$
\Cssi \subseteq \Css \subseteq \Csso\,. 
$$
In particular, if $\Ch$ is degraded, we have 
$$
\Cssi=\Css = \Csso. 
$$
\end{co}

Next, set 
\beqno
{\Cse}\defeq \{(0,R_1,R_{\rm e})\in \Cs \}
\eeqno
and define the secrecy capacity by 
\beqno
C_{\rm ss}(\Ch)\defeq 
 \max_{(R_1,R_1)\in \Cse}R_1
=\max_{(0,R_1)\in \Css}R_1\,.
\eeqno
To describe our result, set
\beqno
& &{\cal R}_{\rm 1e}(U,V,X,S|\Ch)
\\
&\defeq & {\cal R}(U,V,X,S|\Ch)\cap \{(R_0,R_1,R_{\rm e}): R_0=0\}
\\
&=&
\ba[t]{l}
\{(R_1,R_{\rm e}): R_1,R_{\rm e}\geq 0\,,
\vspace{1mm}\\
\ba[t]{rcl}
\:\:R_1 &\leq &I(V;Y|US)
\vspace{1mm}\\
              &     &+ \min \{I(Y;US),I(Z;U|S)\}\,,
\vspace{1mm}\\
\:\:R_{\rm e} & \leq & R_1\,,
\vspace{1mm}\\
\:\:R_{\rm e} &\leq & [I(V;Y|US)-I(V;Z|US)]^{+}\,.\}
\ea
\ea
\eeqno
and 
\beqno
\Csei&\defeq&
\bigcup_{(U,V,X,S)\in {\cal Q}_1}
{\cal R}_{\rm 1e}(U,V,X,S|\Ch)\,,
\\
\Cseo&\defeq&
\bigcup_{(U,V,X,S)\in {\cal Q}_2}
{\cal R}_{\rm 1e}(U,V,X,S|\Ch)\,.
\eeqno
From Theorem \ref{th:sregion}, we have the following corollary.
\begin{co} For any relay channel $\Ch$, 
$$
\Csei \subseteq \Cse \subseteq \Cseo\,. 
$$
Furthermore, 
\beqno
& &\max_{(U,V,X,S)\in {\cal Q}_1}
\left[I(V;Y|US)-I(V;Z|US)\right]^{+}
\\
&\leq &C_{\rm ss}(\Ch)
\\
&\leq & \max_{(U,V,X,S)\in {\cal Q}_2}
\left[
I(V;Y|US)-I(V;Z|US)
\right]^{+}\,.
\eeqno
\end{co}

%In particular, if $\Ch$ is degraded, we have 
%$$
%\Csei=\Cse = \Cseo 
%$$
%and 
%\beqno
%C_{\rm ss}(\Ch)
%&=& \max_{(U,V,X,S)\in {\cal Q}_1}
%\left[I(V;Y|US)-I(V;Z|US)\right]\,.
%\eeqno
%\input{PrDirect.tex}

\section{Gaussian Relay Channels with Confidential Messages}

In this section we study Gaussian relay channels with 
confidential messages, where two channel outputs %of relay channel
are corrupted by additive white Gaussian noises.
Let $(\xi_1, \xi_2)$ be correlated zero mean Gaussian 
random vector with covariance matrix 
$$
\Sigma=\left(
\ba{cc}
N_1& \rho\sqrt{N_1N_2}
\\
\rho\sqrt{N_1N_2} & N_2
\ea
\right)\,, |\rho| <1\,.
$$
Let $\{(\xi_{1,i},\xi_{2,i})\}_{i=1}^{\infty}$ be a sequence 
of independent identically distributed (i.i.d.) zero mean Gaussian 
random vectors. Each $(\xi_{1,i},\xi_{2,i})$ has the covariance 
matrix $\Sigma$. 
The Gaussian relay channel is specified by the above 
covariance matrix $\Sigma$. Two channel outputs $Y_i$ and $Z_i$ 
of the relay channel at the $i$th transmission are give by 
\beqno
Y_i&=&X_i+S_i +\xi_{1,i}\,,
\\
Z_i&=&X_i+\xi_{2,i}\,.
\eeqno
Since $(\xi_{1,i},\xi_{2,i}), i=1,2, \cdots, n$ 
have the covariance matrix $\Sigma$, we have   
$$
\xi_{2,i}= \rho \sqrt{\frac{N_2}{N_1}}\xi_{1,i}+ \xi_{2|1,i}\,,
$$
where $\xi_{2|1,i},i=1,2,\cdots,n $ are zero mean Gaussian 
random variable with variance $(1-\rho^2)N_2$ and 
independent of $\xi_{1,i}$. 
In particular if $\Sigma$ satisfies $N_1 \leq N_2$ 
and $\rho=\sqrt{\frac{N_1}{N_2}}$, we have for 
$i=1,2,$ $\cdots, n$,
%Now, we consider the case where the covariance matrix In this case, we have
\beq
\left.
\ba{l}
Y_i=X_i+ S_i +\xi_{1,i}\,,
\\
Z_i=X_i+ \xi_{1,i}+ \xi_{2|1,i}
\ea
\right\}
\eeq
which implies that for $i=1,2,$ $\cdots, n$,
$Z_i \to (Y_i,S_i) \to X_i$. Hence, the Gaussian relay channel
becomes reversely degraded relay channel.  
Two channel input sequences $\{X_i\}_{i=1}^n$ 
and $\{S_i\}_{i=1}^n$ are subject to 
the following average power constraints:
\beqno
\frac{1}{n}\sum_{i=1}^n{\rm \bf E}\left[X_i^2\right]\leq P_1\,,
\frac{1}{n}\sum_{i=1}^n{\rm \bf E}\left[S_i^2\right]\leq P_2\,.
\eeqno
Let ${\cal R}_{\rm d}(P_1,P_2|\Sigma)$ be a rate region 
for the above Gaussian relay channel when we use a deterministic 
encoder $f_n$. 
To state our result set
\beqno
& & {\cal R}_{\rm d}^{\rm (in)}(P_1,P_2|\Sigma)
\\
&\defeq &
\ba[t]{l}
\{(R_0, R_1, R_{\rm e}): R_0, R_1, R_{\rm e}\geq 0\,,
\vspace{1mm}\\
\ba[t]{rcl}
R_0
&\leq   &{\ds \max_{0\leq \eta \leq 1}}
            \min
        \ba[t]{l}\left\{
           C\left(\frac{\bar{\theta}P_1+P_2
          +2\sqrt{\bar{\theta}\bar{\eta}P_1P_2}}
          {\theta P_1+N_1}
          \right)\right.\,,
          \vspace{1mm}\\
           \quad \! 
        \left.C\left(\frac{\bar{\theta}\eta P_1}{\theta P_1+N_2}\right)
        \right\}\,,
        \ea
        \vspace{1mm}\\
R_1 &\leq &
    C \left(\frac{\theta P_1}{N_1}\right)\,,
\vspace{1mm}\\
R_{\rm e} &\leq & R_1\,,
\vspace{1mm}\\
R_{\rm e}&\leq &
 \left[C \left(\frac{\theta P_1}{N_1}\right)
      -C \left(\frac{\theta P_1}{N_2}\right)\right]^{+}\,,
\vspace{2mm}\\
& & \mbox{ for some }0 \leq \theta \leq 1 \,.\}\,.
\ea
\ea
\eeqno
\beqno
& & {\cal R}_{\rm d}^{\rm (out)}(P_1,P_2|\Sigma)
\\
&\defeq &
\ba[t]{l}
\{(R_0, R_1, R_{\rm e}): R_0, R_1, R_{\rm e}\geq 0\,,
\vspace{1mm}\\
\ba[t]{rcl}
R_0
&\leq & \min
        \ba[t]{l}\left\{
           C\left(\frac{\bar{\theta}P_1+P_2
          +2\sqrt{\bar{\theta}\bar{\eta}P_1P_2}}
          {\theta P_1+N_1}
          \right)\right.\,,
          \vspace{1mm}\\
           \quad \! \left.C\left(\frac{\bar{\theta}\eta P_1}{\theta
	   P_1+N_2}
        \right)
        \right\}\,,
        \ea
        \vspace{1mm}\\
R_1 &\leq &
    C \left(\frac{\theta P_1}
{\frac{(1-\rho^2)N_1N_2 }{ N_1+N_2-2\rho \sqrt{N_1N_2}}}\right)\,,
\vspace{1mm}\\
\:\:R_0+R_1 &\leq & 
       C \left(\frac{P_1+P_2 
         +2\sqrt{\bar{\theta}\bar{\eta}P_1P_2}}{N_1}
          \right)\,,
\vspace{1mm}\\
R_{\rm e} &\leq & R_1\,,
\vspace{1mm}\\
R_{\rm e}&\leq &
 \left[C \left(\frac{\theta P_1}
 {\frac{(1-\rho^2)N_1N_2}{N_1+N_2-2\rho\sqrt{N_1N_2}}}\right)
-C \left(\frac{\theta P_1}{N_2}\right)\right]^{+}\,,
\vspace{2mm}\\
& & \mbox{ for some }0 \leq \theta \leq 1, 0 \leq \eta \leq 1 \,.\}\,.
\ea
\ea
\eeqno
where $C(x)\defeq \frac{1}{2}\log (1+x) \,.$
Our result is the following.
\begin{Th}{\label{th:ThGauss} 
For any Gaussian relay channel, %$\Sigma$, 
\beq
{\cal R}_{\rm d}^{\rm (in)}(P_1,P_2|\Sigma)
\subseteq
{\cal R}_{\rm d}(P_1,P_2|\Sigma)
\subseteq
{\cal R}_{\rm d}^{\rm (out)}(P_1,P_2|\Sigma)\,.
\label{eqn:0aa}
\eeq
In particular, if the relay channel is reversely degraded, i.e., 
$N_1 \leq N_2$ and $\rho=\sqrt{\frac{N_1}{N_2}}$, then
$$
 {}{\cal R}_{\rm d}^{\rm (in)}(P_1,P_2|\Sigma)
={\cal R}_{\rm d}(P_1,P_2|\Sigma)
={}{\cal R}_{\rm d}^{\rm (out)}(P_1,P_2|\Sigma)\,.
$$
}
\end{Th}

Proof of the first inclusion in (\ref{eqn:0aa}) in 
the above theorem is standard. The second inclusion 
can be proved by a converse coding argument similar 
to the one developed by Liang and Veeravalli \cite{lv}. 
Proof of Theorem \ref{th:ThGauss} is stated in the 
next section.   

Next, we study the secrecy capacity of the Gaussian RCCs. 
Define two regions by 
\beqno
& &{\cal C}_{\rm ds}(P_1,P_2|\Sigma)
\nonumber\\
&\defeq &\left\{(R_0,R_1): (R_0,R_1,R_1)
    \in {\cal R}_{\rm d}(P_1,P_2|\Sigma) \right\} \,,
\\
& &{\cal R}_{\rm d1e}(P_1,P_2|\Sigma)
\nonumber\\
&\defeq &\left\{(R_1,R_{\rm e}): (0,R_1,R_{\rm e})
    \in {\cal R}_{\rm d}(P_1,P_2|\Sigma) \right\} \,.
\eeqno
Furthermore, define the secrecy capacity 
$C_{\rm ds}(P_1,P_2|\Sigma)$ by
\beqno
C_{\rm ds}(P_1,P_2|\Sigma)
&\defeq& 
     \max_{(R_1,R_1)\in {\cal R}_{\rm d1e}(P_1,P_2|\Sigma)}R_1
\\
&= & \max_{(0,R_1)\in {\cal C}_{\rm ds}(P_1,P_2|\Sigma)}R_1
\eeqno
We obtain the following two results as corollaries 
of Theorem \ref{th:ThGauss}.
\begin{co}{\rm For any Gaussian relay channel, we have 
$$
{\cal C}_{\rm ds}^{(\rm in)}(P_1,P_2|\Sigma)
\subseteq
{\cal C}_{\rm ds}(P_1,P_2|\Sigma)
\subseteq
{\cal C}_{\rm ds}^{(\rm out)}(P_1,P_2|\Sigma)\,,
$$
where
\beqno
& & {}{\cal C}_{\rm ds}^{\rm (in)}(P_1,P_2|\Sigma)
\\
&\defeq &
\ba[t]{l}
\{(R_0, R_1): R_0, R_1 \geq 0\,,
\vspace{1mm}\\
\ba[t]{rcl}
R_0
&\leq &{\ds \max_{0\leq \eta \leq 1}}
      \min
      \ba[t]{l}\left\{
        C\left(\frac{\bar{\theta}P_1+P_2
        +2\sqrt{\bar{\theta}\bar{\eta}P_1P_2}}
        {\theta P_1+N_1}
        \right)\right.\,,
        \vspace{1mm}\\
        \quad \! 
        \left.C\left(\frac{\bar{\theta}\eta P_1}{\theta P_1+N_2}\right)
        \right\}\,,
        \ea
        \vspace{1mm}\\
R_{1}&\leq &
 \left[C \left(\frac{\theta P_1}{N_1}\right)
 -C \left(\frac{\theta P_1}{N_2}\right)\right]^{+}\,,
\vspace{2mm}\\
& & \mbox{ for some }0 \leq \theta \leq 1 \,.\}\,.
\ea
\ea
\eeqno
\beqno
& & {\cal C}_{\rm ds}^{\rm (out)}(P_1,P_2|\Sigma)
\\
&\defeq &
\ba[t]{l}
\{(R_0, R_1): R_0, R_1\geq 0\,,
\vspace{1mm}\\
\ba[t]{rcl}
R_0
&\leq & {\ds \max_{0\leq \eta\leq 1}}
        \min
        \ba[t]{l}\left\{
          C \left(\frac{\bar{\theta}P_1+P_2
          +2\sqrt{\bar{\theta}\bar{\eta}P_1P_2}}
          {\theta P_1+N_1}
          \right)\right.\,,
          \vspace{1mm}\\
          \quad \! \left.
          C \left(\frac{\bar{\theta}\eta P_1}{\theta P_1+N_2}\right)
         \right\}\,,
         \ea
         \vspace{1mm}\\
R_{1}&\leq &
 \left[C \left(\frac{\theta P_1}
 {\frac{(1-\rho^2)N_1N_2}{N_1+N_2-2\rho\sqrt{N_1N_2}}}\right)
-C \left(\frac{\theta P_1}{N_2}\right)\right]^{+}\,,
\vspace{2mm}\\
& & \mbox{ for some }0 \leq \theta \leq 1\,.\}\,.
\ea
\ea
\eeqno
In particular, if 
$N_1\leq N_2$ and $\rho=\sqrt{\frac{N_1}{N_2}}$,
we have 
$$
 {\cal C}_{\rm ds}^{(\rm in)}(P_1,P_2|\Sigma)
={\cal C}_{\rm ds}(P_1,P_2|\Sigma)
={\cal C}_{\rm ds}^{(\rm out)}(P_1,P_2|\Sigma)\,.
$$
}
\end{co}

\begin{co} For any Gaussian relay channel, we have 
$$
{\cal R}_{\rm d1e}^{(\rm in)}(P_1,P_2|\Sigma)
\subseteq
{\cal R}_{\rm d1e}(P_1,P_2|\Sigma)
\subseteq
{\cal R}_{\rm d1e}^{(\rm out)}(P_1,P_2|\Sigma)\,,
$$
where
\beqno
%& & 
{\cal R}_{\rm d1e}^{\rm (in)}(P_1,P_2|\Sigma)
%\\
&\defeq &
\ba[t]{l}
\{(R_1, R_{\rm e}): R_1, R_{\rm e}\geq 0\,,
\vspace{1mm}\\
\ba[t]{rcl}
R_1 &\leq &
    C \left(\frac{P_1}{N_1}\right)\,,
\vspace{1mm}\\
R_{\rm e} &\leq & R_1\,,
\vspace{1mm}\\
R_{\rm e}&\leq &
 \left[C \left(\frac{P_1}{N_1}\right)
      -C \left(\frac{P_1}{N_2}\right)\right]^{+}\,.\}\,.
\ea
\ea
\eeqno
\beqno
& & {\cal R}_{\rm d1e}^{\rm (out)}(P_1,P_2|\Sigma)
\\
&\defeq &
\ba[t]{l}
\{(R_1, R_{\rm e}): R_1, R_{\rm e}\geq 0\,,
\vspace{1mm}\\
\ba[t]{rcl}
R_1 &\leq &
    C \left(\frac{P_1}
{\frac{(1-\rho^2)N_1N_2 }{ N_1+N_2-2\rho \sqrt{N_1N_2}}}\right)\,,
\vspace{1mm}\\
R_{\rm e} &\leq & R_1\,,
\vspace{1mm}\\
R_{\rm e}&\leq &
 \left[C \left(\frac{P_1}
 {\frac{(1-\rho^2)N_1N_2}{N_1+N_2-2\rho\sqrt{N_1N_2}}}\right)
-C \left(\frac{P_1}{N_2}\right)\right]^{+}\,.\}\,.
\ea
\ea
\eeqno
Furthermore,
\beqno
&   & \ts \left[C \left(\frac{P_1}{N_1}\right)
               -C \left(\frac{ P_1}{N_2}\right)
      \right]^{+}
\nonumber\\
&\leq & C_{\rm ds}(P_1,P_2|\Sigma)
\nonumber\\
&\leq &  \ts \left[C \left(\frac{P_1}
          {\frac{(1-\rho^2)N_1N_2}{N_1
           +N_2-2\rho\sqrt{N_1N_2}}}\right)
           -C \left(\frac{ P_1}{N_2}\right)
     \right]^{+} \,. 
\eeqno
In particular, if $N_1\leq N_2$ and $\rho=\sqrt{\frac{N_1}{N_2}}$,
we have 
$$
{\cal R}_{\rm d1e}^{(\rm in)}(P_1,P_2|\Sigma)
=
{\cal R}_{\rm d1e}(P_1,P_2|\Sigma)
=
{\cal R}_{\rm d1e}^{(\rm out)}(P_1,P_2|\Sigma)\,,
$$
and 
\beqno
C_{\rm ds}(P_1,P_2|\Sigma)
&=& \ts C \left(\frac{P_1}{N_1}\right)
         -C \left(\frac{P_1}{N_2}\right)
        \,.
\eeqno
\end{co}

Note that the secrecy capacity $C_{\rm ds}(P_1,P_2|\Sigma)$ for 
the reversely degraded relay channel does not depend on power 
constraint $P_2$ at the relay. This implies that the security 
of private messages is not affected by the relay.
Leung-Yan-Cheong and Hellman \cite{lh} determined the secrecy capacity 
for the Gaussian wire-tap channel. The above secrecy capacity is equal 
to the secrecy capacity of the Gaussian wire-tap channel derived by them.

\section{Proofs of the Theorems}

In this section we state proofs of Theorems 
\ref{th:ddirect}-\ref{th:ThGauss} stated in the sections III and V. 

In the first subsection we prove Theorem \ref{th:ddirect}, 
the inclusion $\Cdi$ $\subseteq \Cd$ in Theorem 
\ref{th:dregion}, and the inclusion $\Csi$  $\subseteq \Cs$ 
in Theorem \ref{th:sregion}. In the second subsection we prove 
Theorem \ref{th:dconv}, the inclusion $\Cd$ $\subseteq \Cdo$ 
in Theorem \ref{th:dregion}, and the inclusion 
$\Cs$ $\subseteq \Cso$ in Theorem \ref{th:sregion}. 
Proof of Theorem \ref{th:ThGauss} is given in the third subsection.

\subsection{Derivations of the Inner Bounds} 

We first state an important lemma to derive inner bounds. To describe 
this lemma, we need some preparations. Let ${\cal T}_n$, ${\cal J}_n$, 
and ${\cal L}_n$ be three message sets to be transmitted by the sender. 
Let $T_n,$$J_n$, and $L_n$ be uniformly distributed random variable over 
${\cal T}_n$, ${\cal J}_n$ and ${\cal L}_n$ respectively. Elements of 
${\cal T}_n$ and ${\cal J}_n$ are directed to the receiver and relay. 
Elements of ${\cal L}_n$ are only directed to the receiver. Encoder 
function $f_n$ is a one to one mapping from ${\cal T}_n\times$ ${\cal 
J}_n\times$ ${\cal K}_n$ to ${\cal X}^n$. Using the decoder function 
${\psi}_n$, the receiver outputs an element of ${\cal T}_n\times$ ${\cal 
J}_n\times$ ${\cal K}_n$ from a received message of ${\cal Y}^n$. Using 
the decoder function ${\varphi}_n$, the relay outputs an element of 
${\cal T}_n\times$ ${\cal J}_n$ from a received message of ${\cal 
Z}^n$. Formal definitions of ${\psi}_n$ and ${\varphi}_n$ are 
$
 {\psi}_n: {\cal Y}^{n} 
    \to {\cal T}_n \times{\cal J}_n\times{\cal L}_n\,,
{\varphi}_n: {\cal Z}^{n}  \to  {\cal T}_n\times {\cal J}_n\,.
$
We define the average error probability 
of decoding at the receiver over 
${\cal T}_n \times$ 
${\cal J}_n \times$ 
${\cal L}_n$
in the same manner as the definition of $\lambda_1^{(n)}$ 
and use the same notation for this error probability.
We also define the average error probability of decoding 
at the relay over ${\cal T}_n \times$ ${\cal J}_n$ 
in the same manner as the definition of $\lambda_2^{(n)}$ 
and use the same notation for this probability. 
Then, we have the following lemma. 
\begin{lm}\label{lm:direct} \ \ Choose $(U,X,S)$ $\in{\cal P}_1$ 
such that $I(X;Y$ $|YS)$ $\geq$ $I(X;Z|YS)$. Then, there exists 
a sequence of quadruples 
$\{(f_n,\{g_i\}_{i=1}^n,$ 
${\psi }_n ,{\varphi }_n)\}_{n=1}^{\infty}$ 
such that 
\beqno 
   \lim_{n\to\infty}{\lambda}_{1}^{(n)}
&=&\lim_{n\to\infty}{\lambda}_{2}^{(n)}=0\,, 
\\
\lim_{n\to\infty} \nbn \log \pa {\cal T}_n \pa & = & 
\min\{I(Y;US),I(Z;U|S)\}\,,
\\
\lim_{n\to\infty} 
\nbn \log \pa {\cal J}_n \pa &= & I(X;Z|US)\,,
\\
\lim_{n\to\infty} 
\nbn \log \pa {\cal L}_n \pa &= & I(X;Y|US)-I(X;Z|US)\,,
\\
\lim_{n\to\infty} \nbn H(L_n|Z^{n}) &\geq & 
I(X;Y|US)-I(X;Z|US)\,.
\eeqno
\vspace*{1mm}
\end{lm}

The above lemma is proved by a combination of two coding techniques.
One is the method that Csisz\'ar and K\"orner \cite{CsiKor1} used for
deriving an inner bound of the capacity regions of the broadcast
channel with confidential messages and the other is the method that
Cover and El Gamal \cite{cg} developed for deriving a lower bound of
the capacity of the relay channel. Outline of proof of this lemma 
is given in Appendix A. 

\begin{figure}[t]
\bc
\includegraphics[width=7.2cm]{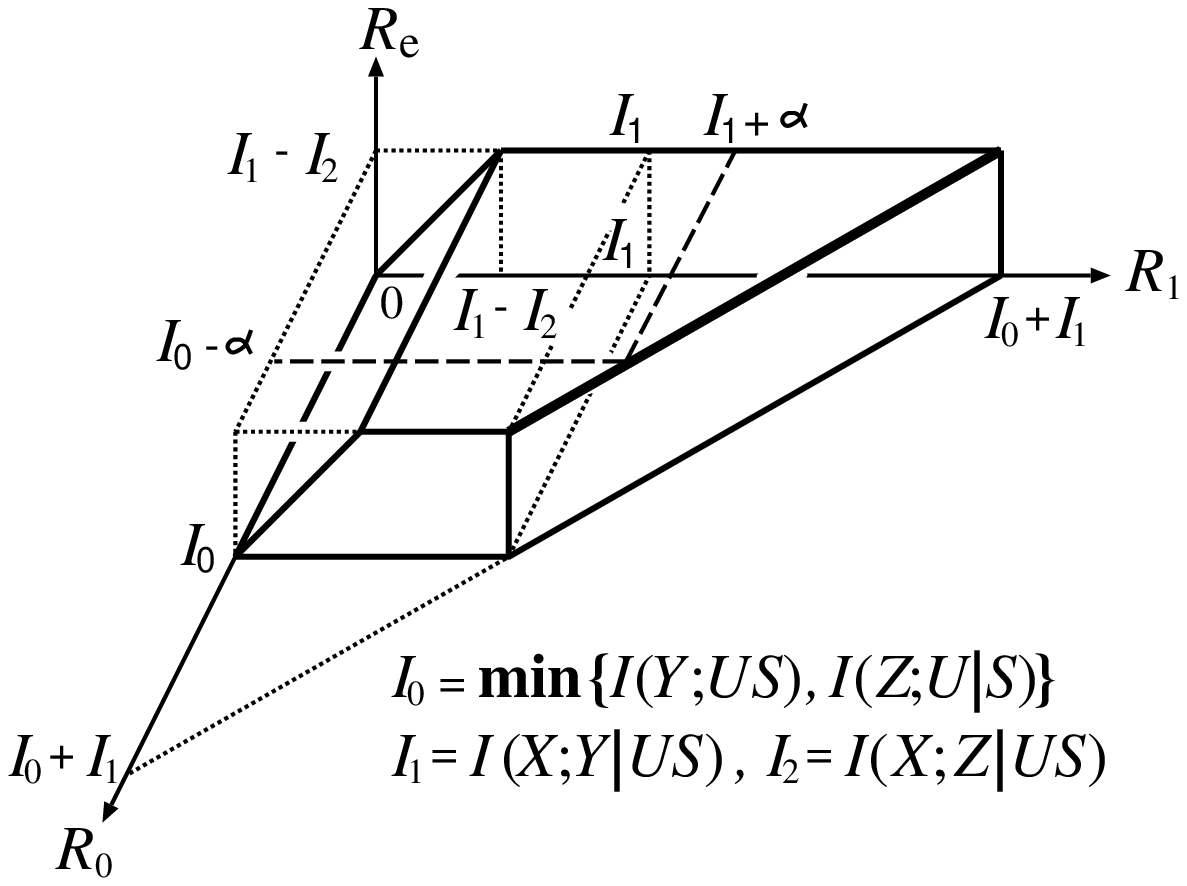}
\caption{
Shape of the region  ${\cal R}(U,X,S|\Ch)$.
} 
\label{fig:FigReg}
\ec
\end{figure}
\begin{figure}[t]
\bc
\includegraphics[width=8.7cm]{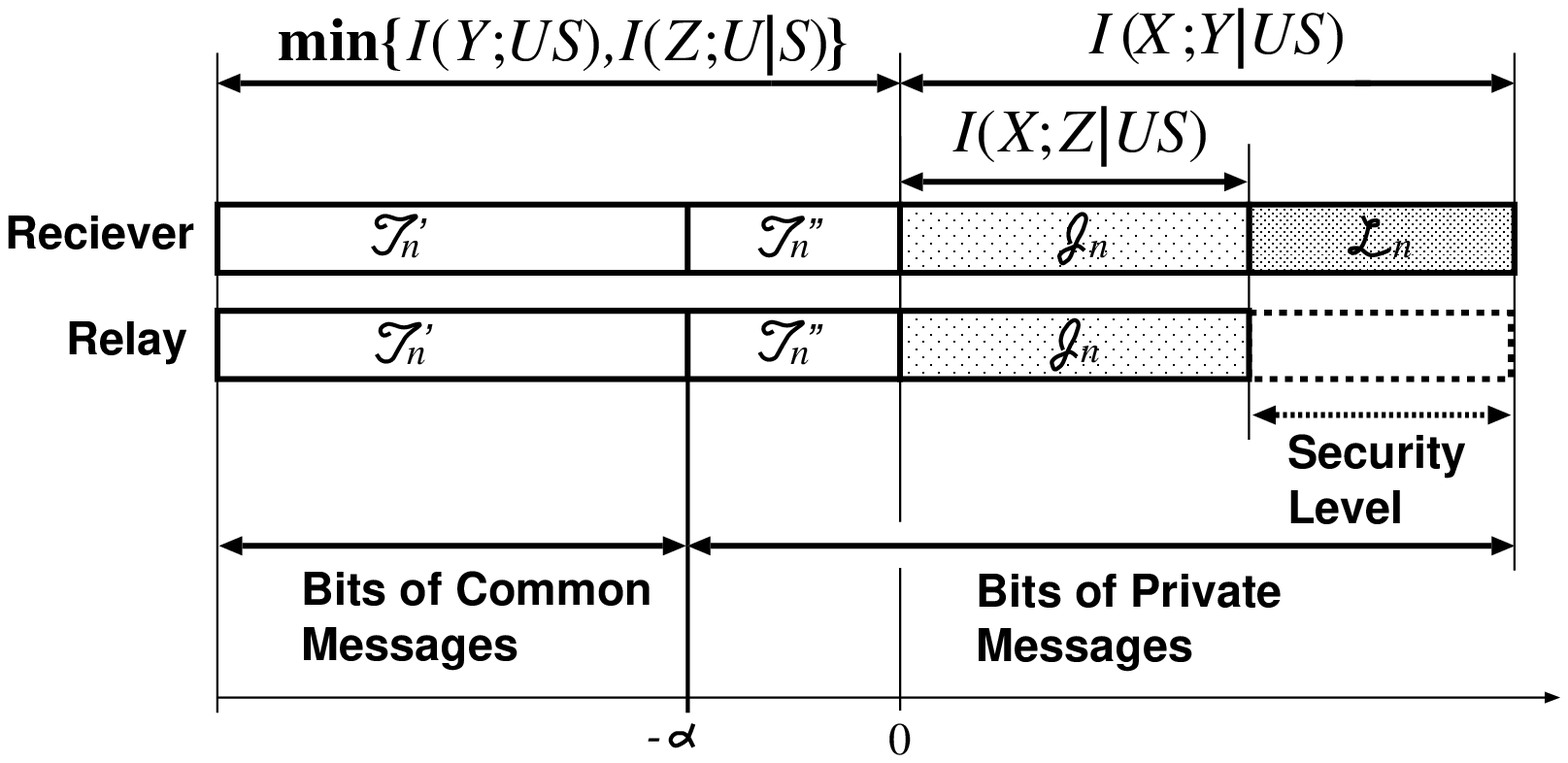}
\caption{
Information contained in the transmitted messages.
} 
\label{fig:info1}
\ec
\vspace*{-5mm}
\end{figure}

{\it Proof of ${\cal R}_{\rm d}^{\rm(in)}(\Ch)\subseteq$ 
              ${\cal R}_{\rm d}(\Ch)$:} \ %The region 
Set
\beqno
I_0&\defeq &\min\{I(Y;US),I(Z;U|S)\}\,,
\\
I_1&\defeq&I(X;Y|US), I_2 \defeq I(X;Z|US)\,.
\eeqno
We consider the case that $I_1\geq I_2$. 
The region ${\cal R}(U,X,S|\Ch)$ in this case 
is depicted in Fig. \ref{fig:FigReg}. 
From the shape of this region 
%the region ${\cal R}(U,X,S|\Ch)$, 
it suffices to show that for every 
$$
\alpha \in [0, \min\{I(Y;US),I(Z;U|S)\}], 
$$
the following $(R_0,R_1,R_{\rm e})$ is achievable: 
\beqno
R_0&=& \min\{I(Y;US),I(Z;U|S)\}-\alpha\,, 
\\
R_1&=& I(X;Y|US)+\alpha\,,
\\
R_{\rm e}&=& I(X;Y|US)-I(X;Z|US)\,. 
\eeqno
%We first consider the case where 
%$$\alpha \in [-\min\{I(Y;US),I(Z;U|S)\},0].
%$$ 
%In this case 
Choose ${\cal T}_n^{\prime}$ 
and ${\cal T}_n^{\prime\prime}$ such that 
\beqno
{\cal T}_n&=&{\cal T}_n^{\prime} 
\times {\cal T}_n^{\prime\prime}\,,
\\
\lim_{n\to\infty}\nbn \log \pa {\cal T}_n^{\prime} \pa 
&=&\min\{I(Y;US),I(Z;U|S)\}-\alpha \,.
\eeqno
We take 
\beqno
{\cal M}_n={\cal T}_n^{\prime}\,,\quad
{\cal K}_n={\cal T}_n^{\prime\prime}
\times{\cal J}_n
\times{\cal L}_n\,.
\eeqno
Then, by Lemma \ref{lm:direct}, we have  
\beqno 
\lim_{n\to\infty}{\lambda}_{1}^{(n)}
&=&\lim_{n\to\infty}{\lambda}_{2}^{(n)}=0, 
\\
\lim_{n\to\infty} \nbn \log \pa {\cal K}_n \pa 
&=& I(X;Y|US)+\alpha\,,
\\
\lim_{n\to\infty} \nbn \log \pa {\cal M}_n \pa 
&=& \min\{I(Y;US),I(Z;U|S)\}-\alpha\,,
\\
\lim_{n\to\infty} \nbn H(K_n|Z^{n}) 
&\geq & 
\lim_{n\to\infty} \nbn H(L_n|Z^{n}) 
\\
&\geq&
I(X;Y|US)-I(X;Z|US)\,.
\eeqno
To help understating the above proof, information 
quantities contained in the transmitted messages are 
shown in Fig. \ref{fig:info1}. 
\hfill\QED
\\

{\it Proof of Theorem \ref{th:ddirect}:} \ 
Since $\tilde{\cal R}_{\rm d}^{\rm(in)}(\Ch)\subseteq $ 
${\cal R}^{ \rm (in) }_{\rm d}(\Ch)$, we have Theorem \ref{th:ddirect}.
\hfill\QED

{\it Proof of $\Csi$ $\subseteq \Cs$:} \ 
%By Lemma \ref{lm:dirZ}, 
%it suffices to show ${\cal R}^{*}_{\rm s}(\Ch)$
%$\subseteq\Cs$. 
%
Choose $(U,V,X,S)\in {\cal Q}_1$. 
The joint distribution of $(U,V,X,S)$ is given by
\beqno  
& & p_{UVXS}(u,v,x,s)
\\
&=&p_{USV}(u,s,v)p_{X|V}(x|v)\,,\:
(u,v,x,s)\in 
{\cal U}\times
{\cal V}\times
{\cal X}\times
{\cal S}\,.
\eeqno
Consider the discrete memoryless channels with input alphabet 
${\cal V}\times{\cal S}$ and output alphabet 
${\cal Y}\times{\cal Z}$, and stochastic 
matrices defined by the conditional 
distribution of $(Y,Z)$ given $V,S$ 
having the form
$$
{\Ch}(y,z|v,s)=\sum_{x\in {\cal X}}{\Ch}(y,z|x,s)
p_{X|V}(x|v)\,.
$$
Any {\it deterministic} encoder 
$f_n^{\prime}:$ ${\cal K}_n\times {\cal M}_n$
$\to {\cal V}^n$ for this new RCC determines 
a {\it stochastic} encoder $f_n$ for the 
original RCC by the matrix product of 
$f_n^{\prime}$ with the stochastic matrix
given by 
$p_{X|V}=$
$\{p_{X|V}(x|v)\}_{(v,x)
\in {\cal V}\times{\cal X}}$. Both encoders 
yield the same stochastic 
connection of messages and received sequences, 
so the assertion follows by applying 
the result of the first inclusion 
%$\Cdi$$\subseteq$$\Cd$
%${\cal R}_{\rm d}^{\rm(in)}(\Ch)\subseteq$ 
%${\cal R}_{\rm d}(\Ch)$
in Theorem \ref{th:dregion} to the new RCC.   
\hfill\QED

Cardinality bounds of auxiliary random variables 
in ${\cal P}_1$ and ${\cal Q}_1$ can be proved by 
the argument that Csisz\'ar and K\"orner \cite{CsiKor1} 
developed in Appendix in their paper.

\subsection{Derivations of the Outer Bounds}

In this subsection we derive the outer bounds stated 
in Theorems \ref{th:dconv}-\ref{th:sregion}. 
We first remark here that cardinality bounds of auxiliary 
random variables in ${\cal P}_2$ and ${\cal Q}_2$ 
in the outer bounds can be proved by the argument 
that Csisz\'ar and K\"orner \cite{CsiKor1} developed 
in Appendix in their paper. 

The following lemma is a basis on derivations of the outer bounds. 
\begin{lm} \label{lm:conv1} 
We assume $(R_0,R_1,R_{\rm e})$ is achievable. Then, we have
\beqno
nR_0&\leq &\min\{I(Y^n;M_n),I(Z^n;M_n)\} +n \delta_{1,n} 
\label{eqn:rate0}\\
nR_1&\leq &I(K_n;Y^n|M_n) +n \delta_{2,n} 
\label{eqn:rate1}\\
n(R_0+R_1)&\leq &I(Y^n;K_nM_n) + n {\delta}_{3,n} 
\label{eqn:rate01}\\
nR_{\rm e}&\leq &nR_1     +n \delta_{4,n} 
\label{eqn:ratee}\\
nR_{\rm e}&\leq &I(K_n;Y^n|M_n)-I(K_n;Z^n|M_n) 
                          +n \delta_{5,n} 
\eeqno
where $\{\delta_{i,n}\}_{n=1}^{\infty}$, $i=1,2,3,4,5$ 
are sequences that tend to zero as $n\to\infty$.
\end{lm}

{\it Proof: }
The above Lemma can be proved by a standard converse 
coding argument using Fano's Lemma. We omit the detail. 
A similar argument is found in Csisz\'ar and K\"orner 
\cite{CsiKor1} in Section V in their paper. 
\hfill\QED

We first prove $\Cd$ $\subseteq$ $\tCdo$. From 
Lemma \ref{lm:conv1}, it suffices to derive upper 
bounds of 
\beqno
& & I(Z^n;M_n), I(Y^n;M_n), I(K_n;Y^n|M_n), 
\\ 
& & I(Y^n;K_nM_n), I(K_n;Y^n|M_n)-I(K_n;Z^n|M_n). 
\eeqno
For upper bound of the above five quantities, we 
have the following Lemma.

\begin{lm} \label{lm:conv2} \ Suppose that $f_n$ is 
a deterministic encoder. Set 
$$U_i\defeq M_nY^{i-1}Z^{i-1}\,,\quad i=1,2, \cdots, n\,.$$
For $i=1,2, \cdots, n$, $U_i$, $X_iS_i$, and $Y_iZ_i$ 
form a Markov chain $U_i$ $\to X_i S_i$ $\to Y_iZ_i$ 
in this order. Furthermore, we have
\beqa
I(Y^n;M_n)&\leq &\sum_{i=1}^{n}I(Y_i;U_iS_i)\,,
\label{eqn:cv1}
\\
I(Z^n;M_n)&\leq &\sum_{i=1}^{n}I(Z_i;U_i|S_i)\,,
\label{eqn:cv2}
\\
I(Y^n;K_nM_n) &\leq &\sum_{i=1}^{n}I(Y_i;X_iS_i)\,,
\label{eqn:cv2xyz}
\\
I(K_n;Y^n|M_n) &\leq &\sum_{i=1}^{n}
I(X_i;Y_iZ_i|U_iS_i)\,,
\label{eqn:cv3}
\eeqa
\beqa
&  &I(K_n;Y^n|M_n)-I(K_n;Z^n|M_n)
\nonumber\\
&\leq &
\sum_{i=1}^{n}I(X_i;Y_i|Z_iU_iS_i)\,. 
\label{eqn:cv4}
\eeqa
\end{lm}

Proof of Lemma \ref{lm:conv2} is given in Appendix B.

{\it Proof of Theorem \ref{th:dconv}:} We assume that 
$(R_0, R_1,R_{\rm e})$ is achievable. 
Let $Q$ be a random variable 
independent of $K_nM_nX^nY^n$ and 
uniformly distributed over 
$\{1,2,\cdots,n\}$.  Set 
\beqa
X \defeq X_Q, S \defeq S_Q, Y \defeq Y_Q, Z\defeq Z_{Q} 
\label{convRv}
\eeqa
Furthermore, set
\beqa
U \defeq U_QQ =Z^{Q-1}Y^{Q-1}M_nQ\,. 
\eeqa
Note that $UXSYZ$ satisfies a Markov chain 
$U\to XS\to YZ$. By Lemmas \ref{lm:conv1} 
and \ref{lm:conv2} we have 
\beq
\left.
\ba{rcl}
R_0 &\leq &\min\{I(Y;US|Q),I(Z;U|SQ)\} +\delta_{1,n} 
\\
    &\leq &\min\{I(Y;US),I(Z;U|S)\} +\delta_{1,n} 
\\
R_1 &\leq &I(X;YZ|US)     +\delta_{2,n} 
\\
R_0+R_1 &\leq & I(XS;Y|Q)   +{\delta}_{3,n} 
\\
        &\leq & I(XS;Y)   +{\delta}_{3,n} 
\\
R_{\rm e}&\leq & R_1     +\delta_{4,n} 
\\
R_{\rm e}&\leq &I(X;Y|ZUS)+\delta_{5,n}\,. 
\ea
\right\}
\label{eqn:conv001}
\eeq
Using memoryless character of the channel it is 
straightforward to verify that $U\to XS\to YZ$ and that 
the conditional distributions of and given coincide 
with the corresponding channel matrix.
Hence by letting $n\to \infty$ in (\ref{eqn:conv001}), we 
obtain $(R_0,R_1,R_{\rm e})\in \tCdo$.
\hfill\QED

Next, we prove the inclusions  
$\Cd$ 
$\subseteq$ 
$\Cdo$ and $\Cd$ $\subseteq$ $\hCdo$.
From Lemma \ref{lm:conv1}, it suffices to derive 
upper bounds of the following five quantities: 
\beqa
& &I(Z^n;M_n),I(Y^n;M_n)\,,
\nonumber\\
& &I(K_n;Y^n|M_n)+I(Y^n;M_n)=I(Y^n;K_nM_n)\,,
\nonumber\\
& &I(K_n;Y^n|M_n)+I(Z^n;M_n)\,,
\label{eqn:conv222}
\\
& &I(K_n;Y^n|M_n)-I(K_n;Z^n|M_n)\,. 
\label{eqn:conv223}
\eeqa
Since
\beqno 
& &I(K_n;Y^n|M_n)+I(Z^n;M_n)
\\
&=&I(K_n;Y^n|M_n)-I(K_n;Z^n|M_n)
+I(K_n M_n;Z^n)\,,
\eeqno
we derive an upper bound of (\ref{eqn:conv222}) 
by estimating upper bounds of $I(K_nM_n;Z^n)$ 
and (\ref{eqn:conv223}).

The following two lemmas are key results to derive the outer bounds.  
\begin{lm}\label{lm:conv3a}\ \ Suppose that $f_n$ is 
a deterministic encoder. Set 
$$
U_i\defeq Y_{i+1}^nZ^{i-1}M_n\,,
\quad i=1,2,\cdots,n \,. 
$$
For $i=1,2, \cdots, n$, $U_i$, $X_iS_iZ_i$, and $Y_i$ 
form a Markov chain $U_i$ $\to X_i Z_iS_i$ $\to Y_i$ 
in this order. Furthermore, we have
\beqa
I(Y^n;M_n)&\leq  &\sum_{i=1}^{n}I(Y_i;U_iS_i)\,,
\\
\label{eqn:cva2x}
I(Z^n;M_n) &\leq &\sum_{i=1}^{n}I(Z_i;U_i|S_i)\,,
\label{eqn:cva1x}
\\
I(Y^n;K_nM_n) &\leq &\sum_{i=1}^{n}I(Y_i;X_iU_iS_i)\,,
\label{eqn:cva22x}
\\
I(Z^n;K_nM_n) &\leq &\sum_{i=1}^{n}I(Z_i;X_iU_i|S_i)\,,
\label{eqn:cva11x}
\eeqa
\beqa
&  &I(K_n;Y^n|M_n{})-I(K_n;Z^n|M_n{})
\nonumber\\
&\leq &
\sum_{i=1}^{n}
\left\{
I(X_i;Y_i|U_iS_i)-I(X_i;Z_i|U_iS_i)
\right\}\,. 
\label{eqn:cva4x}
\eeqa
\end{lm}

\begin{lm}\label{lm:conv3aa} \ \ Suppose that $f_n$ is 
a deterministic encoder. Set 
$$
U_i\defeq Y^{i-1}Z_{i+1}^nM_n\,,\quad i=1,2,\cdots,n\,.
$$
For $i=1,2, \cdots, n$, $U_i$, $X_iS_iZ_i$, and $Y_i$ 
form a Markov 
chain $U_i$ $\to X_i Z_iS_i$ $\to Y_i$ 
in this order. Furthermore, we have
\beqa
I(Y^n;M_n)&\leq &\sum_{i=1}^{n}I(Y_i;U_iS_i)\,,
\\
\label{eqn:cva2z}
I(Z^n;M_n)&\leq &\sum_{i=1}^{n}I(Z_i;U_i|S_i)\,,
\label{eqn:cva1z}
\\
I(Y^n;K_nM_n)&\leq &\sum_{i=1}^{n}I(Y_i;X_iU_iS_i)\,,
\label{eqn:cva22z}
\\
I(Z^n;K_nM_n)&\leq &\sum_{i=1}^{n}I(Z_i;X_iU_i|S_i)\,,
\label{eqn:cva11z}
\eeqa
\beqa
&  &I(K_n;Y^n|M_n{})-I(K_n;Z^n|M_n{})
\nonumber\\
&\leq &
\sum_{i=1}^{n}
\left\{
I(X_iS_i;Y_i|U_i)-I(X_iS_i;Z_i|U_i)
\right. \nonumber\\
& &\left. +I(U_i;Z_i|X_iS_i)\right\}
\nonumber\\
&=&
\sum_{i=1}^{n}
\left\{
I(X_i;Y_i|U_iS_i)-I(X_i;Z_i|U_iS_i)
\right. 
\nonumber\\
& &\left. +\zeta(U_i,S_i,Y_i,Z_i)+I(U_i;Z_i|X_iS_i)\right\}\,. 
\label{eqn:cva4z}
\eeqa
\end{lm}

Proofs of Lemmas \ref{lm:conv3a} and \ref{lm:conv3aa}
are given in Appendixes D and E, respectively. 

{\it Proof of $\Cd$ $\subseteq$ $\Cdo$:} We assume that 
$(R_0,$ $R_1,R_{\rm e})$ is achievable. 
Let $Q$, $X$, $Y$, $Z$, $S$ be the same random 
variables as those in the proof of Theorem \ref{th:dconv}. 
Set  
\beq
U\defeq U_QQ=Y_{Q+1}^nZ^{Q-1}M_nQ\,. 
\label{eqn:convRv2}
\eeq
Note that $UXSYZ$ satisfies a Markov chain $U\to XSZ\to Y$. 
By Lemmas \ref{lm:conv1} and \ref{lm:conv3a}, we have 
\beq
\left.
\ba{rcl}
R_0 &\leq &\min\{I(Y;US),I(Z;U|S)\} +\delta_{1,n} 
\\
R_0+R_1&\leq &I(X;Y|US)
\\
       &     &+\min\{I(Y;US),I(Z;U|S)\}+\tilde{\delta}_{3,n} 
\\
R_{\rm e}&\leq & R_1 +\delta_{4,n} 
\\
R_{\rm e}&\leq &I(X;Y|US)-I(X;Z|US)+\delta_{5,n}\,, 
\ea
\right\}
\label{eqn:conv002}
\eeq
where 
$\tilde{\delta}_{3,n}
\defeq \max\{{\delta}_{1,n}+{\delta}_{2,n}, {\delta}_{3,n}\}\,.$
By letting $n\to \infty$ in (\ref{eqn:conv002}), 
we conclude that $(R_0,R_1,$ $R_{\rm e})$ 
$\in \Cdo$. \hfill\QED

{\it Proof of $\Cd$ $\subseteq$ $\hCdo$:} We assume that 
$(R_0,$ $R_1,R_{\rm e})$ is achievable. 
Let $Q$, $X$, $Y$, $Z$, $S$ be the same random 
variables as those in the proof 
of Theorem \ref{th:dconv}. We set  
\beq
U\defeq U_QQ=Y^{Q-1}Z_{Q+1}^{n}M_nQ\,. 
\label{eqn:convRv22}
\eeq
Note that $UXSYZ$ satisfies a Markov chain 
$U\to XSZ\to Y$.
Furthermore, if $\Ch$ belongs to the independent 
class, we have 
\beq
Z\to XS \to Y\,,\quad U\to XS \to Z,
\label{eqn:Mark}
\eeq
which together with $U\to XSZ\to Y$ yields 
$$
U\to XS \to YZ \,.
$$
By Lemmas \ref{lm:conv1} and \ref{lm:conv3aa}, we have 
\beq
\left.
\ba{rcl}
R_0 &\leq &\min\{I(Y;US),I(Z;U|S)\} +\delta_{1,n} 
\\
R_0+R_1&\leq &I(X;Y|US) + \left[\zeta(U,S,Y,Z)\right]^+
\\
       &     &+\min\{I(Y;US),I(Z;U|S)\}+\tilde{\delta}_{3,n} 
\\
R_{\rm e}&\leq & R_1 +\delta_{4,n} 
\\
R_{\rm e}&\leq &I(XS;Y|U)-I(XS;Z|U)+\delta_{5,n}\,. 
\ea
\right\}
\label{eqn:conv002z}
\eeq
Note here that the quantity $I(U;Z|XS)$ vanishes 
because of the second Markov chain of (\ref{eqn:Mark}).
By letting $n\to \infty$ in (\ref{eqn:conv002z}), 
we conclude that $(R_0,R_1,$ $R_{\rm e})$ 
$\in \hCdo$. \hfill\QED

Finally we prove $\Cs \subseteq	\Cso$. The following is a key 
result to prove the above inclusion.
\begin{lm}\label{lm:conv3b} Suppose that $f_n$ 
is a stochastic encoder. Let $U_i$, $i=1,2,\cdots,n$ 
be the same random variables as those defined 
in Lemma \ref{lm:conv3a}. We further set $V_i \defeq U_iS_iK_n$. 
For $i=1,2,\cdots,n$, $U_iV_iX_iS_iZ_i$ satisfies 
the following Markov chains 
\beqno
& & U_i   \to V_i\to X_iS_iZ_i\to Y_i\,, 
    U_iS_i\to V_iX_i \to Z_i\,, 
\\
& &U_iS_i\to V_i\to X_i\,.
\eeqno
Furthermore, we have
\beqa
I(Y^n;M_n)&\leq &\sum_{i=1}^{n}I(Y_i;U_iS_i)\,,
\label{eqn:cvza1}
\\
I(Z^n;M_n)&\leq &\sum_{i=1}^{n}I(Z_i;U_i|S_i)\,,
\label{eqn:cvza2}
\\
I(Y^n;K_nM_n)&\leq &\sum_{i=1}^{n}I(Y_i;V_iU_iS_i)\,,
\label{eqn:cvza3}
\\
I(Z^n;K_nM_n)&\leq &\sum_{i=1}^{n}I(Z_i;V_iU_i|S_i)\,,
\label{eqn:cvza4}
\eeqa
\beqa
&  &I(K_n;Y^n|M_n{})-I(K_n;Z^n|M_n{})
\nonumber\\
&=&
\sum_{i=1}^{n}
\left\{
I(V_i;Y_i|U_iS_i)-I(V_i;Z_i|U_iS_i)
\right\}\,. 
\label{eqn:cvza5}
\eeqa
\end{lm}
%Combining 
%(\ref{eqn:conv11})-(\ref{eqn:conv13}),(\ref{eqn:conv22}), 
%(\ref{eqn:conv23}), we have 

Proof of Lemma \ref{lm:conv3b} is given in Appendix C.

{\it Proof of $\Cs$ $\subseteq$ $\Cso$: } 
Let $Q$, $X$, $Y$, $Z$, $S$, $U$ be the same 
random variables as those in the proof 
of $\Cd$ $\subseteq$ $\Cdo$. We further set 
$V\defeq USK_n$. Note that $UVXSZ$ satisfies the following 
Markov chains 
\beqno
& & U\to V\to XSZ\to Y\,, 
    US\to VX\to Z\,, 
\\
& &US\to V\to X\,.
\eeqno
By Lemmas \ref{lm:conv1} and \ref{lm:conv3b}
we have 
\beq
\left.
\ba{rcl}
R_0 &\leq &\min\{I(Y;US),I(Z;U|S)\} +\delta_{1,n} 
\\
R_0+R_1&\leq &I(V;Y|US)
\\
       &     &+\min\{I(Y;US),I(Z;U|S)\}+\tilde{\delta}_{3,n} 
\\
R_{\rm e}&\leq & R_1 +\delta_{4,n} 
\\
R_{\rm e}&\leq &I(V;Y|US)-I(V;Z|US)+\delta_{5,n}\,. 
\ea
\right\}
\label{eqn:conv0022}
\eeq
By letting $n\to \infty$ in (\ref{eqn:conv0022}), 
we conclude that $(R_0,R_1,R_{\rm e})$ $\in \Cso$.
\hfill\QED

\subsection{Computation of Inner and Outer Bounds for the Gaussian Relay Channel}

In this subsection we prove Theorem \ref{th:ThGauss}. 
Let $(\xi_1,\xi_2)$ be a zero mean Gaussian random vector with 
covariance $\Sigma$ defined in Section V. By definition, we have 
$$
\xi_2=\rho\sqrt{\frac{N_2}{N_1}}\xi_1
      +\xi_{2|1}\,,
$$
where $\xi_{2|1}$ is a zero mean Gaussian 
random variable with variance $(1-\rho^2)N_2$ and 
independent of $\xi_{1}$. 
We consider the Gaussian relay channel specified by $\Sigma$. 
For two input random variables $X$ and $S$ 
of this Gaussian relay channel, 
output random variables $Y$ and $Z$ are given by
\beqno
Y&=&X+S+\xi_1\,,\\
Z&=&X+\xi_2=X+ \rho\sqrt{\frac{N_2}{N_1}}\xi_1+\xi_{2|1}\,.
\eeqno
Define two sets of random variables by 
\beqno 
{\cal P}(P_1,P_2)
\defeq \{(U,X,S): 
\ba[t]{l}
{\bf E}[X^2]\leq P_1, {\bf E}[S^2]\leq P_2\,,\\
U\to XS\to YZ\: \}  
\ea
\eeqno
\beqno 
{\cal P}_G(P_1,P_2)
\defeq \{(U,X,S): 
\ba[t]{l}
U, X, S\mbox{ are zero mean}\\
\mbox{Gaussian random variables.}
\\
%\mbox{ satisfying}\\ 
{\bf E}[X^2]\leq P_1\,, {\bf E}[S^2]\leq P_2\,,\\
U\to XS\to YZ\:\}\,.  
\ea
\eeqno
Set
\beqno
& &\tilde{\cal R}_{\rm d}^{(\rm out)}(P_1,P_2|\Sigma)
\\
&\defeq &
\ba[t]{l}
\{(R_0,R_1,R_{\rm e}) : R_0,R_1,R_{\rm e} \geq 0\,,
\vspace{1mm}\\
\ba[t]{rcl}
R_0 &\leq & \min \{I(Y;US),I(Z;U|S)\}\,,
\vspace{1mm}\\
R_1 &\leq & I(X;YZ|US)\,,
\vspace{1mm}\\
R_0+R_1 &\leq & I(XS;Y)\,,
\vspace{1mm}\\
R_{\rm e} &\leq &R_1\,,
\vspace{1mm}\\
R_{\rm e} &\leq & I(X;Y|ZUS)\,,
\ea
\vspace{1mm}\\
\mbox{ for some }(U,X,S)\in {\cal P}(P_1,P_2)\,. \}\,.
\ea
\eeqno
\beqno
& &\tilde{\cal R}_{\rm d}^{(\rm in)}(P_1,P_2|\Sigma)
\\
&\defeq &
\ba[t]{l}
\{(R_0,R_1,R_{\rm e}) : R_0,R_1,R_{\rm e} \geq 0\,,
\vspace{1mm}\\
\ba[t]{rcl}
R_0 &\leq & \min \{I(Y;US),I(Z;U|S)\}\,,
\vspace{1mm}\\
R_1 &\leq & I(X;Y|US)\,,
\vspace{1mm}\\
R_{\rm e} &\leq &R_1\,,
\vspace{1mm}\\
R_{\rm e} &\leq & I(X;Y|US)-I(X;Z|US)\,,
\ea
\vspace{1mm}\\
\mbox{ for some }(U,X,S)
\in {\cal P}_G(P_1,P_2)\,. \}\,.
\ea
\eeqno
Then, we have the following.
\begin{Th}{\label{th:GaussConv} For any Gaussian relay channel we have
\beqno
\tilde{\cal R}_{\rm d}^{(\rm in)}(P_1,P_2| \Sigma)
\subseteq 
{\cal R}_{\rm d}(P_1,P_2|\Sigma)
\subseteq 
\tilde{\cal R}_{\rm d}^{(\rm out)}(P_1,P_2| \Sigma)\,.
\eeqno
}\end{Th}

{\it Proof: } The first inclusion can be proved 
by a method quite similar to that in the case 
of discrete memoryless channels. The second inclusion 
can be proved by a method quite similar to 
that in the proof of Theorem \ref{th:dconv}. 
We omit the detail of the proof of those two inclusions.
\hfill\QED 
%the case of discrete memoryless channel.
%e  roof of 
%${\cal R}_{\rm d}^{(\rm in)}(\Ch)$
%$\subseteq {\cal R}_{\rm d}(\Ch)$.

It can be seen from Theorem \ref{th:GaussConv} that 
to prove Theorem \ref{th:ThGauss}, it suffices to prove 
\beqa
& &{\cal R}_{\rm d}^{(\rm in)}(P_1,P_2| \Sigma)
   \subseteq 
   \tilde{\cal R}_{\rm d}^{(\rm in)}(P_1,P_2| \Sigma)
\label{eqn:inc1}\,,\\
& &\tilde{\cal R}_{\rm d}^{(\rm out)}(P_1,P_2| \Sigma)
   \subseteq 
   {\cal R}_{\rm d}^{(\rm out)}(P_1,P_2| \Sigma)\,.
\label{eqn:inc2}
\eeqa
Proof of (\ref{eqn:inc1}) is straightforward. 
To prove (\ref{eqn:inc2}), we need some preparation. 
Set 
\beqno 
a &\defeq &\ts \frac{N_2-\rho\sqrt{N_1N_2}}
              {N_1+N_2-2\rho\sqrt{N_1N_2}}\,.
\eeqno
Define random variables $\tilde{Y}$, $\tilde{\xi}_1$, and $\tilde{\xi}_2$ 
by  
\beqno
\tilde{Y}&\defeq& a Y+\bar{a}Z\,,
\\
\tilde{\xi}_1
& \defeq &a \xi_1 + \bar{a}\xi_2
=\ts \frac{
(1-\rho^2)N_2\xi_1+(N_1-\rho\sqrt{N_1N_2})\xi_{2|1}
 }{N_1+N_2-2\rho\sqrt{N_1N_2}}\,,\\
\tilde{\xi}_2
& \defeq & \xi_1-\xi_{2}=\ts \left(1-\rho \sqrt{\frac{N_2}{N_1}}\right)\xi_1-\xi_{2|1}\,.
\eeqno
Let $\tilde{N}_i={\bf E}[\tilde{\xi}_i^2],i=1,2$. 
Then, by simple computation we can show that $\tilde{\xi}_1$ and $\tilde{\xi}_2$ 
are independent Gaussian random variables and 
\beqno
\tilde{N}_1&=&\ts \frac{(1-\rho^2)N_1N_2}{N_1+N_2-2\rho\sqrt{N_1N_2}}\,,
\\
\tilde{N}_2&=&{N_1+N_2-2\rho\sqrt{N_1N_2}}\,.
\eeqno
We have the following relations between 
$\tilde{Y}$, $Y$, and $Z$: 
\beq
\left.
\ba{rcl}
\tilde{Y}&=& X + aS +\tilde{\xi}_1\,,\\
{Y}&=&\tilde{Y}+ \bar{a}S + \bar{a} \tilde{\xi}_2\,, \\
{Z}&=&\tilde{Y}- a S -a \tilde{\xi}_2\,.
\ea
\right\}
\label{eqn:eqqq}
\eeq
The following is a useful lemma to prove (\ref{eqn:inc2}). 

\begin{lm}\label{lm:Rndau} Suppose that $(U,X,S)$ 
$\in {\cal P}(P_1, P_2)$. Let $X(s)$ be a random 
variable with a conditional distribution of $X$
for given $S=s$. ${\bf E}_{X(s)}[\cdot]$ stands for 
the expectation with respect to the (conditional) 
distribution of $X(s)$. Then, there exists 
a pair $(\alpha,\beta)\in [0,1]^2$ 
such that
\beqno
& &{\bf E}_S\left({\bf E}_{X(S)}X(S)\right)^2=\bar{\alpha} P_1\,,
\\
h(Y|S)&\leq &\ts \frac{1}{2}
\log\left\{(2\pi{\rm e})({\alpha}P_1+N_1)\right\}\,,
\\
h(Z|S)&\leq &\ts \frac{1}{2}
\log\left\{(2\pi{\rm e})({\alpha}P_1+N_2)\right\}\,,
\\
h(Y)&\leq &\ts \frac{1}{2}\log\left\{(2\pi{\rm e})
(P_1+P_2+2\sqrt{\bar{\alpha} P_1P_2}+N_1)\right\}\,,
\\
h(\tilde{Y}|US)&=&\ts \frac{1}{2}
\log \left\{(2\pi {\rm e})(\beta \alpha P_1+ \tilde{N}_1)\right\}\,,
\\ 
h(Y|US)&\geq &\ts \frac{1}{2}
\log \left\{
(2\pi{\rm e})\left(\beta{\alpha}P_1+N_1\right) 
\right\}\,,
\\
h(Z|US)&\geq &\ts \frac{1}{2}
\log \left\{
(2\pi{\rm e})\left(\beta{\alpha}P_1+N_2\right) 
\right\}\,.
\eeqno
\end{lm}

Proof of Lemma \ref{lm:Rndau} is given in Appendix F. 
Using this lemma, we can prove Theorem \ref{th:ThGauss}. 

{\it Proof of Theorem \ref{th:ThGauss}:} \ We first 
prove (\ref{eqn:inc1}). Choose $(U,$ $X,S)\in$ ${\cal P}_G$ 
such that  
\beqno
& &{\bf E}[X^2]=P_1, \quad{\bf E}[S^2]=P_2,
\\
& &U=\ts \sqrt{\frac{\bar{\theta} \bar{\eta} P_1}{P_2}}S 
+ \tilde{U}, \quad X=U+\tilde{X}, 
\eeqno
where $\tilde{U}$ and $\tilde{X}$ are zero mean Gaussian random 
variables with variance $\bar{\theta}\eta P_1$ and $\theta P_1$, 
respectively. The random variables $X$, $S$, $\tilde{U}$, and 
$\tilde{X}$ are independent. For the above choice of 
$(U,X,S)$, we have 
\beqno
I(Y;US)&=& C\left(\ts \frac{\bar{\theta}P_1+P_2
          +2\sqrt{\bar{\theta}\bar{\eta}P_1P_2}}
          {\theta P_1+N_1}
          \right)\,,
\\
I(Z;U|S)&=& C\left(\ts \frac{\bar{\theta}\eta P_1}{\theta P_1+N_2}\right)\,,
\\
I(X;Y|US)&=& C\left(\ts \frac{\theta P_1}{N_1}\right)\,,
\quad
I(X;Z|US)= C\left(\ts \frac{\theta P_1}{N_2}\right)\,.
\eeqno
Thus, (\ref{eqn:inc1}) is proved. Next, we prove 
(\ref{eqn:inc2}). By Lemma \ref{lm:Rndau}, we have 
\beqa
I(Y;US)&=&h(Y)-h(Y|US)
\nonumber\\
      &\leq&C\left(\ts \frac{(1-\beta\alpha)P_1+P_2
            +2\sqrt{ \bar{\alpha}P_1P_2}}
            {\beta \alpha P_1+N_1}
            \right)\,,
\label{eqn:bound1}\\
I(Z;U|S)&=&h(Z|S)-h(Z|US)\nonumber\\
      &\leq& C\left(\ts \frac{\bar{\alpha}P_1}
            {\beta \alpha P_1+N_2}\right)\,,
\label{eqn:bound2}\\
I(XS;Y)&=&h(Y)-h(Y|XS)\nonumber\\
       &\leq& C\left(\ts \frac{(1-\beta \alpha)P_1+P_2
            +2\sqrt{\bar{\alpha}P_1P_2}}{N_1}
        \right)\,,
\label{eqn:bound3}\\
I(X;Z|US)&=& h(Z|US)-h(Z|XS)\nonumber\\
       &\geq & C\left(\ts\frac{\beta \alpha  P_1}{N_2}\right)\,,
\label{eqn:bound31}\\
I(X;YZ|US)&=&h(YZ|US)-h(YZ|XS)\nonumber\\
          &=&h(\tilde{Y}Z|US)-h(\tilde{Y}Z|XS) \nonumber\\
          &=&h(\tilde{Y}|US)+ h(Z|\tilde{Y}US) \nonumber\\
          & &-h(\tilde{Y}|XS)-h(Z|\tilde{Y}XS) \nonumber\\
          &=&h(\tilde{Y}|US)-h(\tilde{Y}|XS)  
\label{eqn:bound3.8}\\
          &=&C \left(\ts \frac{\beta \alpha  P_1}
 {\frac{(1-\rho^2)N_1N_2}{N_1+N_2-2\rho\sqrt{N_1N_2}}}\right)\,, 
\label{eqn:bound4}
\eeqa
where (\ref{eqn:bound3.8}) follows from 
\beqno
%& &
& &h(Z|\tilde{Y}US)=h(Z|\tilde{Y}XS)
=h(Z|\tilde{Y}S)\\
&=&\ts \frac{1}{2}
\log \left\{(2\pi{\rm e}) a^2 \tilde{N_2} \right\}\,. 
\eeqno
From (\ref{eqn:bound31}) and (\ref{eqn:bound4}), we have 
\beqa
%&     &
\hspace*{-4mm}I(X;Y|ZUS)
%\nonumber\\
&\leq & C \left(\ts \frac{\beta \alpha  P_1}
 {\frac{(1-\rho^2)N_1N_2}{N_1+N_2-2\rho\sqrt{N_1N_2}}}\right)
  -C\left(\ts\frac{\beta \alpha  P_1}{N_2}\right)\,.
\label{eqn:bound5}
\eeqa
Here we transform the variable pair 
$(\alpha,\beta)\in [0,1]^2$ into 
$(\eta, \theta)\in [0,1]^2$ in the following manner:
\beq
%\left.
%\ba{rcl}
\theta=\beta\alpha, \quad  
\eta=1-\frac{\bar{\alpha}}{\bar{\theta}}=\frac{\alpha-\theta}{1-\theta}\,. 
\label{eqn:trans0}
\eeq
This map is a bijection because from $(\ref{eqn:trans0})$, 
we have 
\beq
\alpha=1-\bar{\theta}\bar{\eta}\geq \theta, \quad
\beta=\frac{\theta}{\alpha}\,.
\label{eqn:trans1}
\eeq
Combining (\ref{eqn:bound1})-(\ref{eqn:bound3}), (\ref{eqn:bound4}), 
(\ref{eqn:bound5}), and (\ref{eqn:trans1}),
we have (\ref{eqn:inc2}). \hfill\QED 

\section*{\empty}
\appendix

%\subsection{
%}

\subsection{
Outline of Proof of Lemma \ref{lm:direct}
}

Let
\beqno
& &{\cal T}_n=\{1,2,\cdots, 2^{\lfloor nR_0^{(n)}\rfloor}  \}\,,
\quad {\cal L}_n=\{1,2,\cdots,  2^{\lfloor nr_1^{(n)}\rfloor}  \}\,,
\\
& &{\cal J}_n=\{1,2,\cdots, 2^{\lfloor nr_2^{(n)}\rfloor} \}\,,
\eeqno
where $\lfloor x \rfloor$ stands for the integer part of $x$ 
for $x>0$. Furthermore, set
$$
{\cal W}_n\defeq\{1,2,\cdots, 2^{\lfloor nr^{(n)}\rfloor} \}\,.
$$

We consider a transmission over $B$ blocks, each with length $n$. 
For each $i=0,1,\cdots, B-1$, let 
$(w_i,t_i,j_i,l_i)\in 
{\cal W}_n\times 
$$
{\cal T}_n\times 
$$
{\cal J}_n\times 
{\cal L}_n 
$ 
be a quadruple of messages to be transmitted at the $i$th block.
For $i=0$, the constant message vector $(w_0,t_0,j_0,l_0)$ $=(1,1,1,1)$
is transmitted. For fixed $n$, the rate triple
$(%r^{(n)}\frac{B-1}{B},
R_0^{(n)}\frac{B-1}{B}, 
r_1^{(n)}\frac{B-1}{B},r_2^{(n)}\frac{B-1}{B})$        
approaches 
$(%r^{(n)},
R_0^{(n)}, r_1^{(n)}, r_2^{(n)})$ as $B$ $\to \infty$.

We use random codes for the proof. Fix a joint probability distribution 
of $(U,S,X,Y,Z)$:
\beqno
& &p_{USXYZ}(u,s,x,y,z)\\
&=&p_{S}(s)p_{U|S}(u|s)p_{X|US}(x|u,s){\Ch}(y,z|x,s)\,,
\eeqno
where $U$ is an auxiliary random variable that stands for the
information being carried by the message that to be sent 
to the receiver and the relay. In the following, we use 
$A_{\epsilon}$ to denote the jointly $\epsilon$-typical 
set based on this distribution. A formal definition 
of $A_{\varepsilon}$ is in %\cite{CovTh}
[15, Chapter 14.2].      
       
\underline{\bf Random Codebook Generation:} 
We generate a random code book by the following steps.
\begin{itemize} 
\item[1.] Generate $ 2^{\lfloor nr^{(n)}\rfloor}$ 
i.i.d. ${\vc s}\in {\cal S}^n$ 
each with distribution $\prod_{i=1} p_S(s_i).$ 
Index ${\vc s}(w_i), w_i\in {\cal W}_n$. 

\item[2.] For each ${\vc s}(w_i)$, generate 
${2^{\lfloor nR_0^{(n)}\rfloor}}$ i.i.d. 
${\vc u}\in {\cal U}^n$ each with distribution 
$\prod_{i=1} p_U(u_i|s_i)$. Index 
${\vc u}(w_i,t_i), $ $t_i$ $\in$ ${\cal T}_n$.

\item[3.] For each ${\vc u}(t_i,w_i)$ and ${\vc s}(w_i)$, generate 
$2^{ \lfloor n r_1^{(n)}\rfloor }\cdot$ $2^{\lfloor nr_2^{(n)}\rfloor}$ 
i.i.d. ${\vc x}\in {\cal X}^n$ each with distribution  
$\prod_{i=1} $ $p_X(x_i$ $|s_i,u_i)$. Index 
${\vc x}(w_i,t_i,$ $j_i,l_i),$ 
$(w_i,t_i$ $,j_i,$ $l_i)$ 
$\in $ $ {\cal W}_n $ $\times {\cal T}_n$
$\times {\cal J}_n\times {\cal L}_n$.      
\end{itemize}

\underline{\bf Random Partition of Codebook ${\cal T}_n$:}\ \ 
We define the mapping $\phi:{\cal T}_n\to {\cal W}_n$ 
in the following manner. For each $t\in {\cal T}_n$, choose 
$w\in {\cal W}_n$ at random according to the uniform distribution 
over ${\cal W}_n$ and map $t$ to $w$. The random choice 
is independent for each $t\in {\cal T}_n$. For each 
$w\in {\cal W}_n$, define 
${\cal T}_n(w)\defeq$ $\{t\in {\cal T}_n: $ $\phi(t)=w \}\,.$

%For the first block, the message
%In the following ${\sf E}[\cdot]$ stands for the expectation based 
%on the randomness of construction and partition of code books.

\underline{\bf Encoding:}\ \ At the beginning of block $i$, let 
$(t_i,j_i,l_i)$ be the new message triple to be sent 
from the sender in block $i$ and $(t_{i-1},j_{i-1},l_{i-1})$ 
be the message triple to be sent from the sender in 
previous block $i-1$.   

At the beginning of block $i$, the relay has decoded 
the message $t_{i-1}$. It then compute $w_i=\phi(t_{i-1})$ 
and send the codeword ${\vc s}(w_i)$.     

\underline{\bf Decoding:}\ \ 
Let ${\vc y}_i\in {\cal Y}^n$ and 
${\vc z}_i\in {\cal Z}^n$  
be the sequences that the reviver and the relay obtain 
at the end of block $i$, respectively. The decoding 
procedures at the end of block $i$ are as follows. 
%\begin{itemize}

\underline{1. Decoder 2a at the Relay:} \ The relay 
declares that the message $\hat{t}_i$ 
is sent if there is a unique $\hat{t}_i$ such 
that 
$$ 
\left(
{\vc s}(w_i),{\vc u}(w_i,\hat{t}_i), {\vc z}_i
\right)
\in A_{SUZ,\epsilon}\,, 
$$
where $A_{SUZ,\epsilon}$ is a projection of $A_{\epsilon}$ 
along with $(U,S,Z)$, that is
\beqno
A_{SUZ,\epsilon}
&=&\ba[t]{l}
 \{({\vc s},{\vc u},{\vc z}) 
 \in {\cal S}^n \times {\cal U}^n \times {\cal Z}^n: \\
 \quad  ({\vc s},{\vc u},{\vc x},{\vc y},{\vc z})\in A_{\epsilon}\\  
 \quad \mbox{ for some } {\vc x},{\vc y} 
 \in {\cal X}^n\times {\cal Y}^n \}\,. 
 \ea
\eeqno
For projections of $A_\epsilon$, similar definition and notations 
are used for other random variables. It can be shown that 
the decoding error $e_{\rm 2a}^{(n)}$ in this step 
is small for sufficiently large $n$ if 
\beq                    
R_0^{(n)} < I(U;Z|S)\,.
\eeq 

\underline{2. Decoder 2b at the Relay:} \ For 
$(w,t,j)\in {\cal W}_n$ $\times {\cal T}_n $ 
$\times {\cal J}_n $, set
$$
{\cal D}(w,t,j) \defeq \{ {\vc x}: {\vc x}={\vc x}(w,t,j,l) 
\mbox{ for some }l\in {\cal L}_n  \}\,.  
$$
The relay, having known $w_i$ and $\hat{t}_i$, 
declares that the message $\hat{j}_i$ is sent 
if there is a unique $\hat{j}_i$ such that 
\beqno 
& & {\cal D}(w_{i},\hat{t}_{i},\hat{j}_{i})
     \cap 
\\
& &A_{X|SUZ,\epsilon} 
\left({\vc s}(w_i), 
      {\vc u}(w_i,\hat{t}_i),
      {\vc z}_{i-1}
\right)
\neq\emptyset\,,
\eeqno
where
\beqno
& & A_{X|SUZ,\epsilon} 
\left({\vc s}(w_i), 
      {\vc u}(w_i,\hat{t}_i),
      {\vc z}_{i}
\right)
\nonumber\\
&\defeq& 
\{  {\vc x}\in {\cal X}^n : 
    ({\vc s}(w_i),
     {\vc u}(w_i,\hat{t}_i),
     {\vc x},
     {\vc z}_{i}
)\in A_{SUXZ,\epsilon} 
\}\,.
\eeqno
It can be shown that the decoding error $e_{2\rm b}^{(n)}$ 
in this step is small for sufficiently large $n$ if 
$$
r_2^{(n)} < I(X;Z|US)\,. 
$$

\underline{3. Decoders 1a and 1b at the Receiver:} \ The 
receiver declares that the message $\hat{w}_i$ 
is sent if there is a unique $\hat{w}_i$ such that    
$$ 
\left({\vc s}(\hat{w}_i),{\vc y}_i\right)\in A_{SY,\epsilon}\,. 
$$
It can be shown that the decoding error $e_{\rm 1a}^{(n)}$ 
in this step is small for sufficiently large $n$ if  
\beq
r^{(n)}< I(Y;S)\,.
\eeq 
The receiver, having known $w_{i-1}$  and $\hat{w}_i$, 
declares that the message $\hat{\hat{t}}_{i-1}$ is sent 
if there is a unique $\hat{\hat{t}}_{i-1}$ such that 
\beqno
& &\left(
{\vc s}(w_{i-1}),
{\vc u}(w_{i-1}, \hat{\hat{t}}_{i-1}),
{\vc y}_{i-1}
\right)
\in A_{SUY,\epsilon}
\\
& &\mbox{ and } 
\hat{\hat{t}}_{i-1}\in {\cal T}_n(\hat{w}_i). 
\eeqno  
It can be shown that the decoding error $e_{\rm 1b}^{(n)}$ 
in this step is small for sufficiently large $n$ if  
\beqa
R_0^{(n)}
& < & I(Y;U|S) + r^{(n)} 
\nonumber\\
& < &I(Y;U|S)+I(Y;S) =I(Y;US)\,.
\eeqa    

\begin{figure}[t]
\bc
\includegraphics[width=8.8cm]{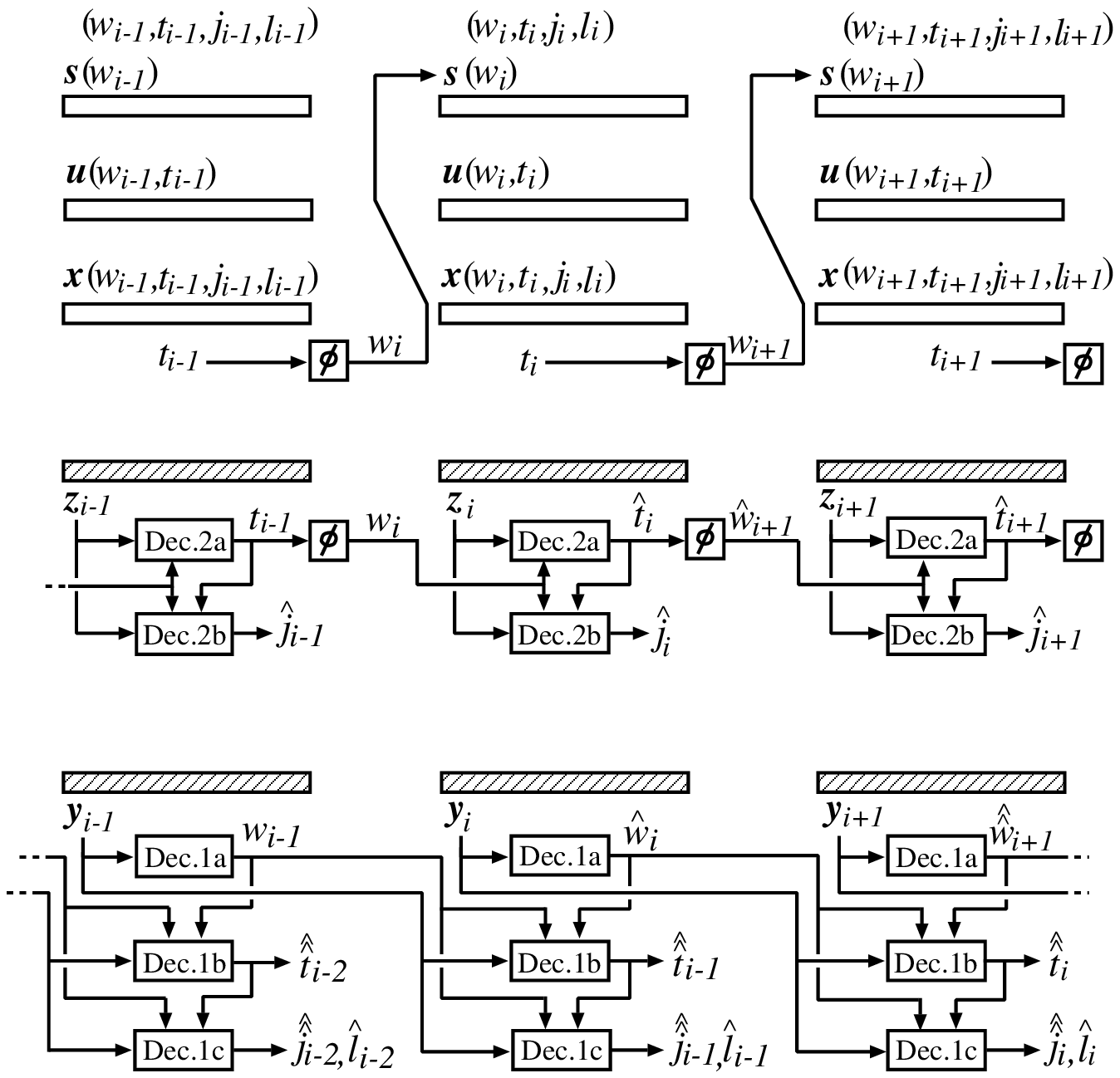}
\caption{
Encoding and decoding processes at 
the blocks $i-1,$ $i,$ and $i+1$.}
\label{fig:cschm} 
\ec
\vspace*{-4mm}
\end{figure}

\underline{4. Decoder 1c at the Receiver:} \ The 
receiver, having known $w_{i-1},$ $\hat{\hat{t}}_{i-1}$, 
declares that the message pair 
$(\hat{\hat{j}}_{i-1},$
$\hat{l}_{i-1})$ is sent 
if there is a unique 
$(\hat{\hat{j}}_{i-1},$
$\hat{l}_{i-1})$ 
such that 
\beqno
& &\left(
{\vc s}(w_{i-1}),
{\vc u}(w_{i-1},\hat{\hat{t}}_{i-1}), 
{\vc x}(w_{i-1},
        \hat{\hat{t}}_{i-1}, 
        \hat{\hat{j}}_{i-1},
        \hat{l}_{i-1}), {\vc y}_{i-1} 
\right)\\
& &\in A_{SUXY,\epsilon}\,.
\eeqno  
It can be shown that the decoding error 
$e_{{\rm 1c}%,i
}^{(n)}$ 
in this step is small for sufficiently large $n$ if  
\beq
r_2^{(n)}+r_1^{(n)}< I(X;Y|US)\,.
\eeq    
For convenience we show the encoding and decoding processes 
at the blocks $i-1,$ $i,$ and $i+1$ in Fig. \ref{fig:cschm}.       
Summarizing the above argument, 
it can be shown that for each block $i=1,2,\cdots, B-1$, 
there exists a sequence of code books 
\beqno
& &\left\{({\vc s}(w_i),{\vc u}(w_i,t_i), \right. \\
& &\quad \left. {\vc x}(w_i,t_i,j_i,l_i)\right\}_{(w_i,t_i,j_i,l_i)
           \in{\cal W}_n\times {\cal T}_n
       \times {\cal J}_n\times {\cal L}_n}
\eeqno
for $n=1,2,\cdots, $ such that
\beqa
    \lim_{n\to\infty}\frac{1}{n} \log |{\cal T}_n| 
&=& \lim_{n\to\infty}R_0^{(n)} = \min\{I(Y;US), I(Z;U|S)\}
\nonumber\\
    \lim_{n\to\infty}\frac{1}{n} \log |{\cal J}_n| 
&=& \lim_{n\to\infty}r_2^{(n)} = I(X;Z|US)
\label{eqn:zzp0} \\
    \lim_{n\to\infty}\frac{1}{n} \log |{\cal L}_n| 
&=& 
\lim_{n\to\infty}[r_1^{(n)}+r_2^{(n)}]-\lim_{n\to\infty}r_2^{(n)}
\nonumber\\
&=&I(X;Y|US)-I(X;Z|US)
\label{eqn:zzp} 
\eeqa
\beqa
    \lim_{n\to\infty}e_{\rm 1b}^{(n)} 
&=& \lim_{n\to\infty}e_{\rm 1c}^{(n)} =0\,,
\nonumber\\
    \lim_{n\to\infty}e_{\rm 2a}^{(n)} 
&=& \lim_{n\to\infty}e_{\rm 2b}^{(n)} =0\,.
\label{eqn:zss}
\eeqa

\underline{\bf Computation of Security Level:} \ 
Suppose that $T_n, L_n, J_n$ are random variables 
corresponding the messages to be transmitted 
at the block $i$. For simplicity of notations we 
omit the suffix $i$ indicating the block number in those 
random variables. For each block $i=1,2,\cdots, B-1$, 
we estimate a lower bound of $H(L_n|Z^n)$. Let ${{\Rw}_n}$ 
be a random variable over ${\cal W}_n$ induced by $\phi$ 
and the uniform random variable $\tilde{T}_n$ 
over ${\cal T}_n$ corresponding to the messages to be transmitted 
at the block $i-1$. Formally, $W_n=\phi(\tilde{T}_n)$.
On lower bound of $H(L_n|Z^n)$, 
we have the following :   
\beqa
& & H(L_n|Z^n)=H(L_nJ_n|Z^n)-H(J_n|Z^nL_n)
\nonumber\\
&\geq & H(J_nL_n|Z^nT_n{\Rw}_n)-H(J_n|Z^n)
\label{eqn:zz00}
\eeqa
By Fano's inequality, we have   
\beq
\frac{1}{n}H(J_n|Z^n) \leq r_2^{(n)} e_{\rm 2b}^{(n)} + \frac{1}{n}\,. 
\label{eqn:zza}
\eeq
The right member of (\ref{eqn:zza}) tends to zero 
as $n\to\infty$. Hence, it suffices to evaluate a lower 
bound of $H(L_nJ_n|Z^nT_n{\Rw}_n)$. 
On this lower bound we have the following 
chain of inequalities: 
\beqa
& &H(J_nL_n|Z^n{\Rw}_n T_n)
\nonumber\\
&=& H(J_nL_nZ^n|{\Rw}_nT_n)-H(Z^n|{\Rw}_nT_n) 
\nonumber\\
&=& H(J_nL_n|{\Rw}_nT_n)
\nonumber\\
& &+H(Z^n|{\Rw}_nT_nJ_nL_n)-H(Z^n|{\Rw}_nT_n)
\nonumber\\
&=& \log\left\{ |{\cal J}_n| |{\cal L}_n|\right\}
\nonumber\\
& &+H(Z^n|{\Rw}_nT_nJ_nL_n)-H(Z^n|{\Rw}_nT_n)
\nonumber\\
&\geq& n[r_2^{(n)}+r_1^{(n)}]-2
\nonumber\\ 
& & +H(Z^n|{\Rw}_nT_nJ_nL_n)-H(Z^n|{\Rw}_nT_n)\,.
\label{eqn:za}
\eeqa
We first estimate $H(Z^n|{\Rw}_nT_nJ_nL_n)$. To this end we set  
\beqno
{\cal A}^{*}
&=&
\ba[t]{l}\left\{( w,t,j,l,{\vc z}): \right.\\
\quad \left. ({\vc s}(w),{\vc x}(w,t,j,l),{\vc z}) 
\in A_{SXZ,\epsilon}\right\} \\ 
\ea
\eeqno
By definition of ${\cal A}^*$, if $(w,t,j,l,{\vc z})$
$\in {\cal A}^*$, we have 
\beqno
& &\left|
   -\frac{1}{n}\log p_{Z^n|X^nS^n}
  ({\vc z}|{\vc x}(w,t,j,l),{\vc s}(w))
\right.\\
& &\qquad \quad\biggl. -H(Z|XS) \biggr|
\leq 2\epsilon\,. 
\eeqno
Then, we have  
\beqa
&     &  H(Z^n|{\Rw}_nT_nJ_nL_n)
\nonumber\\
&\geq & n[H(Z|XS)-2\epsilon]
\Pr\{ ( {\Rw}_n,T_n,J_n,L_n, Z^n )\in {\cal A}^{*} \} 
\nonumber\\
&\geq & n[H(Z|XS)-2\epsilon](1-e_{2b}^{(n)})\,.
\label{eqn:aa0}
\eeqa
%Hence we have 
Next, we derive an upper bound of $H(Z^n|{\Rw}_n T_n)$. 
To this end we set  
\beqno
{\cal B}^{*}
=\left\{
(w,t,{\vc z}): ({\vc s}(w),{\vc u}(w,t),{\vc z}) 
\in A_{SUZ,\epsilon}
\right\}
\eeqno
By definition of ${\cal B}^*$, if 
$(w,t,{\vc z})$
$\in {\cal B}^*$, we have      
\beqno
& &\left| -\frac{1}{n}
\log p_{Z^n|{U}^nS^n}({\vc z}|{\vc u}(w,t),{\vc s}(w))
-nH(Z|US) \right|
\leq 2\epsilon\,. 
\eeqno
Then, we have  
\beqa
&     &  H(Z^n|{\Rw}_nT_n)
\nonumber\\
&\leq & n[H(Z|US)+2\epsilon]+n \kappa \Pr\{({\Rw}_n,T_n, Z^n )\notin {\cal B}^{*} \} 
\nonumber\\
&\leq & n[H(Z|US)+2\epsilon]+n\kappa e_{\rm 2a}^{(n)}\,,
\label{eqn:aabbb}
\eeqa
where $\kappa=\max_{(s,u,z)}\log[ p_{Z|US}(z|u,s)^{-1}]\,.$
Combining (\ref{eqn:zz00})
%, (\ref{eqn:za}), (\ref{eqn:aa0}), 
%and 
-(\ref{eqn:aabbb}), we have 
\beqa
&     &\frac{1}{n}H(L_n|Z^n)
\nonumber\\
& \geq & r_2^{(n)}+ r_1^{(n)}-I(X;Z|US)
\nonumber\\
&  & -4\epsilon-\frac{3}{n} - \kappa e_{\rm 2a}^{(n)}
     -[r_2^{(n)}+H(Z|XS)]e_{\rm 2b}^{(n)}\,.
\label{eqn:zzz0z}
\eeqa
From 
(\ref{eqn:zzp0})
%, 
% (\ref{eqn:zzp}), 
-(\ref{eqn:zss}), and  (\ref{eqn:zzz0z}), we have
\beqno
& &\lim_{n \to \infty}\frac{1}{n}H(L_n|Z^n)
\geq 
 I(X;Y|US)-I(X;Z|US)-4\epsilon\,.
\eeqno
Since $\epsilon$ can be made arbitrary small, we have
\beqno
& &\lim_{n\to \infty}\frac{1}{n}H(L_n|Z^n)\geq 
 I(X;Y|US)-I(X;Z|US)\,.
\eeqno
For $n=1,2,\cdots$, we choose block $B=B_n$ so that 
\beq
B_n=\left\lfloor \left(\max\{
e_{\rm 1b}^{(n)},
e_{\rm 1c}^{(n)},
e_{\rm 2a}^{(n)},
e_{\rm 2b}^{(n)}\}\right)^{-1/2} \right\rfloor\,.
\eeq
Define $\{g_i\}_{i=1}^{nB_n}$ by 
$$
g_i \defeq 
\left\{
\ba{ll} 
\phi,  &\mbox{ if }i\mbox{ mod }n=0\,,\\ 
\mbox{constant}, &\mbox{ otherwise}\,. 
\ea
\right.
$$
Then, we obtain the desired result for a sequence of block codes 
$
\left\{
(f_{nB_n},\{g_i\}_{i=1}^{nB_n}, \psi_{nB_n},\varphi_{nB_n})
\right\}_{n=1}^{\infty}
$\,.
Thus, the proof of Lemma \ref{lm:direct} is 
completed. 
\hfill\QED

\subsection{
Proof of Lemma \ref{lm:conv2}
}

In the following bounding argument we frequently use equalities 
or data processing inequalities based on the fact that for 
$i=1,2,\cdots, n$, $S_i=g_i($ $Z^{i-1})$ 
is a function of $Z^{i-1}$. The notation $[i]$ 
stands for $\{1,2,\cdots,n\}$$-\{i\}$.

{\it Proof of Lemma \ref{lm:conv2}:} We first prove (\ref{eqn:cv1}).  
We have the following chain of inequalities:
\beqa
& &I(Y^n;M_n)
\nonumber\\
&=&H(Y^n)-H(Y^n|M_n)
\nonumber\\
&=&\sum_{i=1}^n
\left\{H(Y_i|Y^{i-1})-H(Y_i|Y^{i-1}M_n)\right\}
\nonumber\\
&\leq &\sum_{i=1}^n
\left\{H(Y_i)-H(Y_i|Y^{i-1}Z^{i-1}M_n)\right\}
\nonumber\\
&=&\sum_{i=1}^n
   \left\{H(Y_i)-H(Y_i|Y^{i-1}Z^{i-1}S_{i}M_n)
    \right\}
\nonumber\\
&=&\sum_{i=1}^n I(Y_i;U_iS_i)\,.
\nonumber
\eeqa
Next, we  prove (\ref{eqn:cv2}). 
We have the following chain of inequalities:
\beqa
& &I(Z^n;M_n)
\nonumber\\
&=&H(Z^n)-H(Z^n|M_n)
\nonumber\\
&=&\sum_{i=1}^n
\left\{H(Z_i|Z^{i-1})-H(Z_i|Z^{i-1}M_n)\right\}
\nonumber\\
&\leq &\sum_{i=1}^n
 \left\{H(Z_i|S_i)-H(Z_i|Y^{i-1}Z^{i-1}S_iM_n)\right\}
\nonumber\\
&=&\sum_{i=1}^n I(Z_i;U_i|S_i)\,.
\nonumber
\eeqa
Thirdly, we prove (\ref{eqn:cv2xyz}). 
We have the following chain of inequalities:
\beqa
& &I(Y^n;K_nM_n)
\nonumber\\
&=&\sum_{i=1}^n
\left\{H(Y_i|Y^{i-1})-H(Y_i|Y^{i-1}K_nM_n)\right\}
\nonumber\\
&=&\sum_{i=1}^n
\left\{H(Y_i|Y^{i-1})-H(Y_i|Y^{i-1}X^n)\right\}
\label{eqn:conv8z8}\\
&\leq &\sum_{i=1}^n
\left\{H(Y_i)-H(Y_i|Y^{i-1}S_iX^n)\right\}
\nonumber\\
&= &\sum_{i=1}^n
\left\{H(Y_i)-H(Y_i|X_iS_i)\right\}
\label{eqn:conv8z9}\\
&=&\sum_{i=1}^n I(Y_i;X_iS_i)\,,
\nonumber
\eeqa
where\\
(\ref{eqn:conv8z8}): $X^n=f_n(K_n,M_n)$ 
and $f_n$ is a one-to-one mapping.\\
(\ref{eqn:conv8z9}): 
$Y_i\to X_iS_i \to Y^{i-1}X_{[i]}\,.$\\
Next, we prove (\ref{eqn:cv3}). 
We have the following chain of inequalities:
\beqa
& &I(K_n;Y^n|M_n)
\nonumber\\
&\leq&I(K_n;Y^nZ^n|M_n)
\nonumber\\
&=& H(Y^nZ^n|M_n)-H(Y^nZ^n|K_nM_n)
\nonumber\\
&=& H(Y^nZ^n|M_n)-H(Y^nZ^n|X^nM_n)
\label{eqn:conv888}\\
&=&\sum_{i=1}^n
\left\{H(Y_iZ_i|Y^{i-1}Z^{i-1}M_n)\right.
\nonumber\\
& & \left.\qquad\quad 
-H(Y_iZ_i|Y^{i-1}Z^{i-1}S_iX^nM_n)\right\}
\nonumber\\
&=&\sum_{i=1}^n \left\{H(Y_iZ_i|U_iS_i)
-H(Y_iZ_i|U_iS_iX^n)\right\}
\nonumber\\
&=&\sum_{i=1}^n \left\{H(Y_iZ_i|U_iS_i)
-H(Y_iZ_i|X_iS_i)\right\}
\label{eqn:conv889}\\
&\leq&\sum_{i=1}^n \left\{H(Y_iZ_i|U_iS_i)-H(Y_iZ_i|U_iX_iS_i)\right\}
\nonumber\\
&=&\sum_{i=1}^n I(X_i;Y_iZ_i|U_iS_i)\,,
\nonumber
\eeqa
where\\
(\ref{eqn:conv888}): $X^n=f_n(K_n,M_n)$ and $f_n$ is a one-to-one mapping.\\
(\ref{eqn:conv889}): $Y_iZ_i\to X_iS_i \to U_iX_{[i]}\,.$\\
Finally, we prove (\ref{eqn:cv4}). 
We have the following chain of inequalities:
\beqa
& &I(K_n;Y^n|M_n)-I(K_n;Z^n|M_n)
\nonumber\\
&\leq&I(K_n;Y^nZ^n|M_n)-I(K_n;Z^n|M_n)
\nonumber\\
&= &I(K_n;Y^n|Z^nM_n)
\nonumber\\
&= &H(Y^n|Z^nM_n)-H(Y^n|Z^nK_nM_n)
\nonumber\\
&=&H(Y^n|Z^nM_n)-H(Y^n|Z^nX^n)
\label{eqn:conv990}\\
&=&\sum_{i=1}^n
\left\{H(Y_i|Y^{i-1}Z^{n}M_n)-H(Y_i|Y^{i-1}Z^{n}X^n)\right\}
\nonumber\\
&=  &\sum_{i=1}^n\left\{H(Y_i|U_iS_iZ_i^n)-H(Y_i|Y^{i-1}S_iZ^nX^n)\right\}
\nonumber\\
&\leq&\sum_{i=1}^n\left\{H(Y_i|U_iS_iZ_i)-H(Y_i|Y^{i-1}S_iZ^nX^n)\right\}
\nonumber\\
&=&  \sum_{i=1}^n\left\{H(Y_i|U_iS_iZ_i)-H(Y_i|S_iZ_iX_i)\right\}
\label{eqn:conv991}\\
&\leq&\sum_{i=1}^n\left\{H(Y_i|U_iS_iZ_i)-H(Y_i|U_iS_iZ_iX_i)\right\}
\nonumber\\
&=   &\sum_{i=1}^n I(X_i;Y_i|Z_iU_iS_i)\,,
%\nonumber\\
%&=   &\sum_{i=1}^n\left\{I(X_i;Y_iZ_i|U_iS_i)-I(X_i;Z_i|U_iS_i)\right\}\,,
\nonumber
\eeqa
where \\
(\ref{eqn:conv990}): $X^n=f_n(K_n,M_n)$ and $f_n$ is a one-to-one mapping.\\
(\ref{eqn:conv991}):  $Y_i\to Z_iX_iS_i \to Y^{i-1}Z_{[i]}^nX_{[i]}\,.$\\
Thus, the proof of Lemma \ref{lm:conv2} is completed.
\hfill\QED

\subsection{
Proof of Lemma \ref{lm:conv3b}
}

%We first present a useful identity to prove those lemmas.
%in  subsequent arguments. %to derive outer bounds.
%\begin{lm}(Csisz\'ar and K\"orner \cite{CsiKor1}) \label{lm:conv2.9}
%\beqno
%& &I(K_n;Y^n|M_nS^n)-I(K_n;Z^n|M_nS^n)
%\nonumber\\
%& & I(Y^n;K_n|M_nS^n)-I(Z^n;K_n|M_nS^n)
%\nonumber\\
%&=  & \sum_{i=1}^n
%        \left\{
%         I(K_n;Y_i|Y^{i-1}Z_{i+1}^nM_nS^n)
%       \right.
%\nonumber\\
%&     &\qquad\left. 
%        -I(K_n;Z_i|Y^{i-1}Z_{i+1}^nM_nS^n)
%        \right\}\,.
%\label{eqn:conv228}
%\eeqno
%\end{lm}
The following is a key lemma to prove Lemma \ref{lm:conv3b}.
\begin{lm} \label{lm:conv3d}
\beqa
I(Y^n;M_n)
&\leq&
\sum_{i=1}^nI(Y_i;Y_{i+1}^nZ^{i-1}M_nS_i)\,,
\label{eqn:cvk1a}\\ 
I(Z^n;M_n) 
&\leq&\sum_{i=1}^nI(Z_i;Y_{i+1}^nZ^{i-1}M_n|S_i)\,,
\label{eqn:cvk2a}
\eeqa
\beqa 
&    &I(K_n;Y^n|M_n) + I(Y^n;M_n)
\nonumber\\ 
&\leq&\sum_{i=1}^n
      \left\{
       I(K_n;Y_i|Y_{i+1}^nZ^{i-1}M_nS_i)\right.
\nonumber\\
&  & \qquad \left. 
        +I(Y_i;Y_{i+1}^nZ^{i-1}M_nS_i)
      \right\}\,,
\label{eqn:cvk3}\\ 
&     &I(K_n;Y^n|M_n)+I(Z^n;M_n)
\nonumber\\
&\leq&\sum_{i=1}^n
       \left\{
       I(K_n;Y_i|Y_{i+1}^nZ^{i-1}M_nS_i)\right.
\nonumber\\
& &    \qquad \left. 
       +I(Z_i;Y_{i+1}^nZ^{i-1}M_n|S_i)
       \right\}\,,
\label{eqn:cvk4}\\
& & I(Y^n;K_n|M_n)-I(Z^n;K_n|M_n)
\nonumber\\
&=  & \sum_{i=1}^n
        \left\{
        I(K_n;Y_i|Y_{i+1}^nZ^{i-1}M_nS_i)
       \right.
\nonumber\\
&     &\qquad\left. 
        -I(K_n;Z_i|Y_{i+1}^nZ^{i-1}M_nS_i)
        \right\}\,.
\label{eqn:cvk5}
\eeqa
\end{lm}

Lemma \ref{lm:conv3b} immediately follows from 
the above lemma. We omit the detail. In the remaining part 
of this appendix we prove Lemma \ref{lm:conv3d}.

{\it Proof of Lemma \ref{lm:conv3d}:}
We first prove (\ref{eqn:cvk1a}) and (\ref{eqn:cvk2a}). 
We have the following chains of inequalities:  
\beqa
& & I(Y^n;M_n)
\nonumber\\
&=&\sum_{i=1}^n
\left\{
H(Y_i|Y_{i+1}^n)-H(Y_i|Y_{i+1}^nM_n)
\right\}
\nonumber\\
&\leq&\sum_{i=1}^n
\left\{
H(Y_i)-H(Y_i|Y_{i+1}^nZ^{i-1}S_i M_n)
\right\}
\nonumber\\
&=&\sum_{i=1}^n
I(Y_i;Y_{i+1}^nZ^{i-1}S_iM_n)\,,
\nonumber\\
& & I(Z^n;M_n)
\nonumber\\
&=&\sum_{i=1}^n
\left\{
H(Z_i|Z_{i-1})-H(Z_i|Z_{i-1}M_n)
\right\}
\nonumber\\
&\leq&\sum_{i=1}^n
\left\{
H(Z_i|S_i)-H(Z_i|Y_{i+1}^nZ^{i-1}S_iM_n)
\right\}
\nonumber\\
&=&\sum_{i=1}^n
I(Z_i;Y_{i+1}^nZ^{i-1}M_n|S_i)\,.
\nonumber
\eeqa
Next, we prove (\ref{eqn:cvk3}). We have the following 
chain of inequalities: 
\beqa
& & I(K_n;Y^n|M_n)+I(Y^n;M_n)
\nonumber\\
&=& H(Y^n)-H(Y^n|K_nM_n)
\nonumber\\
&=&\sum_{i=1}^n
\left\{
H(Y_i|Y_{i+1}^n)-H(Y_i|Y_{i+1}^nK_nM_n)
\right\}
\nonumber\\
&\leq&\sum_{i=1}^n
\left\{
H(Y_i)-H(Y_i|Y_{i+1}^nZ^{i-1}S_iK_nM_n)
\right\}
\nonumber\\
&=&\sum_{i=1}^n I(Y_i;Y_{i+1}^nZ^{i-1}S_iK_nM_n)
\nonumber\\
&=&\sum_{i=1}^n 
\left\{
I(Y_i;K_n|Y_{i+1}^nZ^{i-1}S_iM_n) 
\right.
\nonumber\\
& &\qquad \left. + I(Y_i;Y_{i+1}^nZ^{i-1}S_iM_n) 
\right\}\,.
\nonumber
\eeqa
Finally, we prove (\ref{eqn:cvk4}) and (\ref{eqn:cvk5}). 
We first observe the following two identities. 
\beqa
\hspace*{-4mm}& & H(Y^n|M_n) - H(Z^n|M_n)
\nonumber\\
\hspace*{-4mm}&=&\sum_{i=1}^n
\left\{ 
H(Y_i|Y_{i+1}^n Z^{i-1} M_n) - H(Z_i|Y_{i+1}^n Z^{i-1}M_n)
\right\}\,,
\label{eqn:cvk61} 
\\
\hspace*{-4mm}& & H(Y^n|K_nM_n) - H(Z^n|K_nM_n)
\nonumber\\
\hspace*{-4mm}&=&\sum_{i=1}^n
\left\{ 
H(Y_i|Y_{i+1}^n Z^{i-1} K_nM_n)
\right.
\nonumber\\
\hspace*{-4mm}& &\qquad\left. 
-H(Z_i|Y_{i+1}^n Z^{i-1}K_nM_n)
\right\}
\label{eqn:cvk70}\,. 
\eeqa
Those identities follow from an elementary computation based 
on the chain rule of entropy. 
The equality of (\ref{eqn:cvk5}) immediately follows 
from $(\ref{eqn:cvk61})-(\ref{eqn:cvk70})\,$. Now 
we proceed to the proof of (\ref{eqn:cvk4}).
We have the following chains of inequalities:  
\beqa
& & I(K_n;Y^n|M_n)+I(Z^n;M_n)
\nonumber\\
&=& H(Y^n|M_n)-H(Y^n|K_nM_n)+H(Z^n)-H(Z^n|M_n)
\nonumber\\
&=&\sum_{i=1}^n
   \left\{
       H(Y_i|Y_{i+1}^nZ^{i-1}M_n) -H(Z_i|Y_{i+1}^nZ^{i-1}M_n)
   \right\}
\nonumber\\
& &+\sum_{i=1}^n
    \left\{
     H(Z_i|Z^{i-1})
    -H(Y_i|Y_{i+1}^nK_nM_n)
    \right\}
\nonumber\\
&\leq &\sum_{i=1}^n
   \left\{
    H(Y_i|Y_{i+1}^nZ^{i-1}S_iM_n)
   -H(Z_i|Y_{i+1}^nZ^{i-1}S_iM_n)
   \right\}
\nonumber\\
& &+\sum_{i=1}^n
    \left\{
     H(Z_i|S_i)
    -H(Y_i|Y_{i+1}^nZ^{i-1}S_iK_nM_n)
    \right\}
\nonumber\\
&=&\sum_{i=1}^n
   \left\{
    I(K_n;Y_i|Y_{i+1}^nZ^{i-1}S_iM_n)
   \right.
\nonumber\\
& &\qquad
+\left.
    I(Z_i;Y_{i+1}^nZ^{i-1}M_n|S_i)
   \right\}\,.
\nonumber
\eeqa
Thus, the proof of Lemma \ref{lm:conv3d} is completed.\hfill\QED  
%We postpone the proof of this lemma to the end 
%of this subsection.

\subsection{
Proof of Lemma \ref{lm:conv3a} %and \ref{lm:conv3aa}
}

In this appendix we prove Lemma \ref{lm:conv3a}. %and \ref{lm:conv3aa}.
We first present a lemma necessary to prove this lemma.
\begin{lm}\label{lm:convlmD}\ \ Suppose that $f_n$ is a deterministic encoder. 
Set $X^n=f_n(K_nM_n)$. For any sequence $\{U_i\}_{i=1}^{n}$ 
of random variables, we have 
\beqa 
& &I(Y^n;K_nM_n) \leq \sum_{i=1}^nI(Y_i;X_iU_iS_i)
\label{eqn:convZ1}
\\
& &I(Z^n;K_nM_n) \leq \sum_{i=1}^nI(Z_i;X_iU_i|S_i)
\label{eqn:convZ2}
\eeqa
\end{lm}

{\it Proof:} We first prove (\ref{eqn:convZ1}). We have the 
following chain of inequalities:   
\beqa
& &I(Y^n;K_nM_n)
\nonumber\\
&=& H(Y^n)-H(Y^n|K_nM_n)
\nonumber\\
&=& H(Y^n)-H(Y^n|X^n)
\label{eqn:convZ3}\\
&=    & \sum_{i=1}^n \left\{ H(Y_i|Y^{i-1})-H(Y_i|Y^{i-1}X^n)\right\}
\nonumber\\
&\leq & \sum_{i=1}^n \left\{ H(Y_i)-H(Y_i|Y^{i-1}X^nS_i)\right\} 
\nonumber\\
&=&  \sum_{i=1}^n \left\{H(Y_i)-H(Y_i|X_iS_i)\right\} 
\label{eqn:convZ4}\\
&\leq&  \sum_{i=1}^n \left\{H(Y_i)-H(Y_i|X_iU_iS_i)\right\} 
\nonumber\\
&=& \sum_{i=1}^n I(Y_i;X_iU_iS_i)\,,
\nonumber
\eeqa
where \\
(\ref{eqn:convZ3}): $X^n=f_n(K_n,M_n)$ and $f_n$ is a one-to-one mapping.\\
(\ref{eqn:convZ4}): $Y_i\to X_iS_i \to Y^{i-1}X_{[i]}\,.$\\
Next, we prove (\ref{eqn:convZ2}). We have the 
following chain of inequalities:   
\beqa
& &I(Z^n;K_nM_n)
\nonumber\\
&=& H(Z^n)-H(Z^n|K_nM_n)
\nonumber\\
&=& H(Z^n)-H(Z^n|X^n)
\label{eqn:convZ5}\\
&=    & \sum_{i=1}^n \left\{ H(Z_i|Z^{i-1})-H(Z_i|Z^{i-1}X^n)\right\}
\nonumber\\
&=    & \sum_{i=1}^n \left\{ H(Z_i|Z^{i-1})-H(Z_i|Z^{i-1}X^nS_i)\right\} 
\nonumber\\
&\leq & \sum_{i=1}^n \left\{ H(Z_i|S_i)-H(Z_i|Z^{i-1}X^nS_i)\right\}
\nonumber\\
&=    & \sum_{i=1}^n \left\{ H(Z_i|S_i)-H(Z_iX_iS_i)\right\}
\label{eqn:convZ6}\\
&\leq & \sum_{i=1}^n \left\{ H(Z_i|S_i)-H(Y_i|X_iU_iS_i)\right\}
\nonumber\\
&=    & \sum_{i=1}^nI(Z_i;X_iU_i|S_i)\,,
\nonumber
\eeqa
where \\
(\ref{eqn:convZ5}): $X^n=f_n(K_n,M_n)$ and $f_n$ is a one-to-one mapping.\\
(\ref{eqn:convZ6}): $Z_i\to X_iS_i \to Z^{i-1}X_{[i]}\,.$\\
Thus, the proof of Lemma \ref{lm:convlmD} is completed.
\hfill\QED
% in  subsequent arguments. %to derive outer bounds.

{\it Proof of Lemma \ref{lm:conv3a}:} 
%Then, we have  
Set $U_i=Y_{i+1}^nZ^{i-1}M_n$. It can easily be verified that $U_i$, 
$X_iS_iZ_i$, $Y_i$ form a Markov chain 
$U_i \to X_iS_iZ_i \to Y_i $
in this order.  
From (\ref{eqn:cvk1a}), (\ref{eqn:cvk2a}), and (\ref{eqn:cvk5}) 
in Lemma \ref{lm:conv3d}, we obtain   
\newcommand{\wide}{\hspace*{-8mm}}
\beqno
I(Y^n;M_n)&\leq& \sum_{i=1}^nI(Y_i;U_iS_i),
\\ 
I(Z^n;M_n)&\leq& \sum_{i=1}^nI(Z_i;U_i|S_i), 
\eeqno
and 
\beqa
&     &I(Y^n;K_n|M_n)-I(Z^n;K_n|M_n)
\nonumber\\
&\leq & 
      \sum_{i=1}^n\{I(K_n;Y_i U_iS_i)
                   -I(K_n;Z_i U_iS_i)\}\,,
\label{eqn:conv13}
\eeqa
respectively. 
From (\ref{eqn:convZ1}), (\ref{eqn:convZ2}) 
in Lemma \ref{lm:convlmD}, we obtain   
\beqno
I(Y^n;K_nM_n)&\leq& \sum_{i=1}^nI(Y_i;X_iU_iS_i),
\\ 
I(Z^n;K_nM_n)&\leq& \sum_{i=1}^nI(Z_i;X_iU_i|S_i), 
\eeqno
respectively. 
It remains to evaluate an upper bound of 
$$
I(K_n;Y_i U_iS_i)-I(K_n;Z_i U_iS_i)\,.
$$
We have the following chain of inequalities:
\beqa
& &  I(K_n;Y_i U_iS_i)-I(K_n;Z_i U_iS_i)
\nonumber\\
&=& H(Y_i|U_iS_i)-H(Y_i|K_nM_nU_iS_i)
\nonumber\\
& &-H(Z_i|U_iS_i)+H(Z_i|K_nM_nU_iS_i)
\nonumber\\
&= & H(Y_i|U_iS_i)-H(Y_i|X^nU_iS_i)
\nonumber\\
&  &-H(Z_i|U_iS_i)+H(Z_i|X^nU_iS_i)
\label{eqn:conv999}\\
&= & H(Y_i|U_iS_i)
\nonumber\\
&  &-H(Y_i|Z_iX^nU_iS_i)-I(Y_i;Z_i|X^nU_iS_i)
\nonumber\\
&  &-H(Z_i|U_iS_i)
\nonumber\\
&  &+H(Z_i|Y_iX^nU_iS_i)+I(Y_i;Z_i|X^nU_iS_i)
\nonumber\\
&= & H(Y_i|U_iS_i)-H(Y_i|Z_iX^nU_iS_i)
\nonumber\\
&  &-H(Z_i|U_iS_i)+H(Z_i|Y_iX^nU_iS_i)
\nonumber\\
&= & H(Y_i|U_iS_i)-H(Y_i|Z_iX_iS_i)
\nonumber\\
&  &-H(Z_i|U_iS_i)+H(Z_i|Y_iX^nU_iS_i)
\label{eqn:conv1000}\\
&\leq & H(Y_i|U_iS_i)-H(Y_i|Z_iX_iU_iS_i)
\nonumber\\
&     &-H(Z_i|U_iS_i)+H(Z_i|Y_iX_iU_iS_i)
\nonumber\\
&=& I(Y_i;Z_iX_i|U_iS_i)-I(Z_i;Y_iX_i|U_iS_i)
\nonumber\\
&=& I(X_i;Y_i|U_iS_i)-I(X_i;Z_i|U_iS_i)\,,
\label{eqn:conv23}
\nonumber
\eeqa
where \\
(\ref{eqn:conv999}):  $X^n=f_n(K_n,M_n)$ and $f_n$ is a one-to-one mapping.\\
(\ref{eqn:conv1000}): $Y_i\to Z_iX_iS_i \to U_iX_{[i]}\,.$\\
Thus, the proof of Lemma \ref{lm:conv3a} is completed. \hfill\QED  

\subsection{
Proof of Lemma \ref{lm:conv3aa} %and \ref{lm:conv3aa}
}

In this appendix we prove Lemma \ref{lm:conv3aa}.

{\it Proof of Lemma \ref{lm:conv3aa}:} \ \  
Set $U_i\defeq Y^{i-1}Z_{i+1}^{n}M_n$. 
It can easily be verified that $U_i$, 
$X_iS_iZ_i$, $Y_i$ form a Markov chain 
$U_i \to X_iS_iZ_i \to Y_i $
in this order.  
In a manner similar to the proof of Lemma \ref{lm:conv3d}, 
we can derive the following two bounds  
\beqa
I(Y^n;M_n)
&\leq&
\sum_{i=1}^nI(Y_i;Y^{i-1}Z_{i+1}^{n}M_nS_i)\,,
\label{eqn:cvk1}\\ 
I(Z^n;M_n) 
&\leq&\sum_{i=1}^nI(Z_i;Y^{i-1}Z_{i+1}^{n}M_n|S_i)\,.
\label{eqn:cvk2}
\eeqa
Hence, we have 
\beqno
I(Y^n;M_n)&\leq& \sum_{i=1}^nI(Y_i;U_iS_i),
\\ 
I(Z^n;M_n)&\leq& \sum_{i=1}^nI(Z_i;U_i|S_i). 
\eeqno
Furthermore, from (\ref{eqn:convZ1}), (\ref{eqn:convZ2}) 
in Lemma \ref{lm:convlmD}, we obtain   
\beqno
I(Y^n;K_nM_n)&\leq& \sum_{i=1}^nI(Y_i;X_iU_iS_i),
\\ 
I(Z^n;K_nM_n)&\leq& \sum_{i=1}^nI(Z_i;X_iU_i|S_i), 
\eeqno
respectively. 
It remains to evaluate an upper bound of 
$$
I(K_n;Y^n|M_n)-I(K_n;Z^n|M_n)\,.
$$
Since $f_n$ is a deterministic, we have
\beqa
& &I(K_n;Y^n|M_n)-I(K_n;Z^n|M_n)\,
\nonumber\\
&=&H(Y^n|M_n)-H(Z^n|M_n)
   -H(Y^n|X^n)
\nonumber\\
& &+H(Z^n|X^n)\,. 
\label{eqn:conv777} 
\eeqa
We separately evaluate the following two quantities:
\beqno
&&H(Y^n|M_n)-H(Z^n|M_n)\,,
\\
&&H(Y^n|X^n)-H(Z^n|X^n)\,.
\eeqno
We observe the following two identities: 
\beqa
\hspace*{-9mm}& & H(Y^n|M_n) - H(Z^n|M_n)
\nonumber\\
\hspace*{-9mm}&=&\sum_{i=1}^n
\left\{ 
H(Y_i|Y^{i-1} Z_{i+1}^n M_n)-H(Z_i|Y^{i-1}Z_{i+1}^nM_n)
\right\},\label{eqn:cvk61b} 
\\
\hspace*{-9mm}& & -H(Y^n|X^n) + H(Z^n|X^n)
\nonumber\\
\hspace*{-9mm}&=&\sum_{i=1}^n
\left\{ 
-H(Y_i|Y_{i+1}^n Z^{i-1} X^n)
+H(Z_i|Y_{i+1}^n Z^{i-1}X^n)
\right\}. 
\label{eqn:cvk70b}
\eeqa
Those identities follow from an elementary computation based 
on the chain rule of entropy. 
From (\ref{eqn:cvk61b}), we have  
\beqa
& &H(Y^n|M_n)-H(Z^n|M_n)
\nonumber\\
&=&\sum_{i=1}^n \left\{H(Y_i|U_i)-H(Z_i|U_i)\right\}\,.
\label{eqn:cvk62b} 
\eeqa
Next, we evaluate an upper bound of     
$$
-H(Y_i|Y_{i+1}^n Z^{i-1} X^n)
+H(Z_i|Y_{i+1}^n Z^{i-1}X^n)\,.
$$
Set 
$
\tilde{U}_i\defeq Y_{i+1}^n Z^{i-1}X_{[i]}\,.
$
We have the following chain of inequalities:  
\beqa 
& & -H(Y_i|Y_{i+1}^n Z^{i-1} X^n)
    +H(Z_i|Y_{i+1}^n Z^{i-1}X^n)
\nonumber\\
&=&-H(Y_i|X_i\tilde{U}_i)+H(Z_i|X_i\tilde{U}_i)
\nonumber\\
&= &-H(Y_i|X_iS_i\tilde{U}_i)+H(Z_i|X_iS_i\tilde{U}_i)
\nonumber\\
&= &-H(Y_i| Z_iX_iS_i\tilde{U_i})
    +I(Y_i;Z_i|X_iS_i\tilde{U_i})
\nonumber\\
&  &+H(Z_i| Y_iX_iS_i\tilde{U_i})
    -I(Y_i;Z_i|X_iS_i\tilde{U_i})
\nonumber\\
&= &-H(Y_i|Z_iX_iS_i\tilde{U_i})+ H(Z_i|Y_iX_iS_i\tilde{U_i})
\nonumber\\
&=    &-H(Y_i|Z_iX_iS_i) +H(Z_i|Y_iX_iS_i\tilde{U}_i)
\label{eqn:conv1001}\\
&\leq &-H(Y_i|Z_iX_iS_i) +H(Z_i|Y_iX_iS_i)
\nonumber\\
& =   &   -H(Y_i|X_iS_i)+I(Y_i;Z_i|X_iS_i)
\nonumber\\
&     &   +H(Z_i|X_iS_i)-I(Y_i;Z_i|X_iS_i)
\nonumber\\
&=    &   -H(Y_i|X_iS_i) +H(Z_i|X_iS_i)\,,
\label{eqn:conv23b}
\eeqa
where (\ref{eqn:conv1001}) follows 
from $Y_i\to$ $Z_iX_iS_i$ $\to \tilde{U}_i\,.$
Combining 
(\ref{eqn:conv777}), 
(\ref{eqn:cvk70b}), 
(\ref{eqn:cvk62b}), 
and (\ref{eqn:conv23b}), 
we obtain
\beqno
& &I(K_n;Y^n|M_n)-I(K_n;Z^n|M_n)
\nonumber\\
&\leq &\sum_{i=1}^n \{H(Y_i|U_i) -H(Z_i|U_i)
\nonumber\\
& &\qquad      -H(Y_i|X_iS_i)+H(Z_i|X_iS_i)\}
\nonumber\\
&\leq &\sum_{i=1}^n \{H(Y_i|U_i) -H(Z_i|U_i)
\nonumber\\
& &\qquad        -H(Y_i|X_iS_iU_i)+H(Z_i|X_iS_i)\}
\nonumber\\
&= &\sum_{i=1}^n\{I(X_iS_i;Y_i|U_i)-I(X_iS_i;Z_i|U_i)
\nonumber\\
& &\qquad+I(U_i;Z_i|X_iS_i)\}
\nonumber\\
&= &\sum_{i=1}^n\{I(X_i;Y_i|U_iS_i)-I(X_i;Z_i|U_iS_i)
\nonumber\\
& &\qquad+\zeta(S_i,U_i,Y_i,Z_i)+I(U_i;Z_i|X_iS_i)\}\,.
\eeqno
Thus, the proof of Lemma \ref{lm:conv3aa} is completed.
\hfill\QED
%{\it Proof of 
%\input{apd1.tex}
%\input{apd2.tex}

\subsection{
Proof of Lemma \ref{lm:Rndau} %and \ref{lm:conv3aa}
}

We first observe that by the Cauchy-Schwarz inequality we have 
\beqno
{\bf E}_S\left({\bf E}_{X(S)}X(S)\right)^2
&\leq &{\bf E}_S
\left[
\left(
\sqrt{{\bf E}_{X(S)} X^2(S) }\sqrt{ {\bf E}_{X(S)} 1} 
\right)^2
\right]\\
&= & {\bf E}_S{\bf E}_{X(S)}X^2(S)\leq P_1\,.
\eeqno
Then, there exits $\alpha \in [0,1]$ such that
\beqno
&&{\bf E}_S\left({\bf E}_{X(S)}X(S)\right)^2=\bar{\alpha} P_1\,.
\eeqno
We derive an upper bound of $h(Y)$. We have the following 
chain of inequalities:   
\beqa
& &h(Y)
\nonumber\\
&=&h(X+S+\xi_1)
\nonumber\\
&\leq &{\ts \frac{1}{2}} 
\log
\left\{
(2\pi{\rm e})\left({\bf E}_{XS}|X+S|^2+ N_1\right)
\right\}
\nonumber\\
&=&{\ts \frac{1}{2}} 
   \log
   \left\{
   (2\pi{\rm e})
   \left({\bf E}_{X}|X|^2
       +2{\bf E}_{XS}XS 
        +{\bf E}_{S}S^2 +N_1 \right)
   \right\}
\nonumber\\
&\leq& {\ts \frac{1}{2}} 
   \log
   \left\{
   (2\pi{\rm e})
   \left(P_1+P_2+2{\bf E}_{XS}XS 
        +N_1 \right)
   \right\}\,.
\label{eqn:ppp0} 
\eeqa
By the Cauchy-Schwarz inequality we have 
\beqa
&    &{\bf E}_{XS}XS
\nonumber\\ 
& =  &{\bf E}_S\left[S{\bf E}_{X(S)}X(S)\right]
\nonumber\\
&\leq&\sqrt{{\bf E}_SS^2}\sqrt{{\bf E}_{S}\left({\bf E}_{X(S)}X(S)\right)^2}
 =\sqrt{P_2}\sqrt{\bar{\alpha}P_1}\,.
\label{eqn:ppp} 
\eeqa
From (\ref{eqn:ppp0}) and (\ref{eqn:ppp}),
we have 
$$
h(Y)\leq\ts 
{\ts \frac{1}{2}} \log
\left\{ (2 \pi{\rm e})\left(P_1+P_2+\sqrt{\bar{\alpha}P_1 P_2}+N_1\right)
\right\}\,.
$$
Next, we estimate an upper bound of $h(Y|S)$. 
We have the following chain of inequalities:
\beqa
h(Y|S)&=&{\bf E}_S\left[h(X(S)+\xi_1)\right]
\nonumber\\ 
      &\leq&{\bf E}_S
\left[\ts \frac{1}{2}
\log\left\{(2\pi{\rm e})
\left({\bf V}_{X(S)}\left[X(S)\right]+ N_1 \right)
     \right\}
\right]
\nonumber\\
&=&{\bf E}_S
\Bigl[\ts \frac{1}{2}
\log
 \Bigl\{(2\pi{\rm e})
 \Bigl(
       {\bf E}_{X(S)}[X^2(S)]
 \Bigr.\Bigr.\Bigr.
\nonumber\\
& &\left.
    \left.
     \left.
   -\left({\bf E}_{X(S)}X(S)\right)^2+ N_1 
     \right)
   \right\}
  \right]
\nonumber\\
&\leq & \ts \frac{1}{2}
\log
 \Bigl\{(2\pi{\rm e})
 \Bigl(
       {\bf E}_S{\bf E}_{X(S)}[ X^2(S)]
 \Bigr.\Bigr.
\nonumber\\
& &\left.
   \left. 
   -{\bf E}_S\left({\bf E}_{X(S)}X(S)\right)^2+ N_1 
  \right)
 \right\}
\nonumber\\
&\leq & \ts \frac{1}{2}\log
   \left\{(2\pi{\rm e})\left(\alpha P_1+ N_1 \right)\right\}\,.
\nonumber
\eeqa
Similarly, we obtain 
\beqa
h(Z|S)&\leq& \ts \frac{1}{2}\log
   \left\{(2\pi{\rm e})\left(\alpha P_1+ N_2 \right)\right\}\,,
\nonumber\\
h(\tilde{Y}|S)&\leq& \ts \frac{1}{2}\log
   \left\{(2\pi{\rm e})\left(\alpha P_1+ \tilde{N}_1\right)\right\}\,.
\label{eqn:zzz0}
\eeqa
Since 
\beqno
&& h(\tilde{Y}|S) \geq h(\tilde{Y}|XS)
=\ts \frac{1}{2}\log \left\{(2\pi{\rm e})\tilde{N}_1\right\}
\eeqno
and (\ref{eqn:zzz0}), there exists $\beta\in [0,1]$ such that
$$
h(\tilde{Y}|US)=
\ts \frac{1}{2}\log
\left\{(2\pi{\rm e})\left(\beta\alpha P_1+ \tilde{N}_1\right)\right\}\,.
$$
Finally, we derive lower bounds of $h(Y|US)$ and $h(Z|US)$. We recall  
the following relations between $Y,Z,$ and $\tilde{Y}$:
\beqa
Y&=&\tilde{Y}+\bar{a}S+\bar{a}\tilde{\xi_2}\,,
\label{eqn:zzoo}\\
Z&=&\tilde{Y}-{a}S-{a}\tilde{\xi_2}\,.
\label{eqn:zzooo}
\eeqa
Applying entropy power inequality to (\ref{eqn:zzoo}), 
we have
\beqno  
\ts \frac{1}{2 \pi{\rm e}} 2^{2h(Y|US)}
&\geq& \ts \frac{1}{2\pi{\rm e}} 2^{2h(\tilde{Y}|US) }
          +\frac{1}{2\pi{\rm e}} 2^{2h(\bar{a}{\tilde{\xi}_2})} 
\nonumber\\
&=& \beta\alpha P_1 + \tilde{N}_1 + \bar{a}^2\tilde{N}_2
\nonumber\\
&=& \beta\alpha P_1+\ts \frac{(1-\rho^2)N_1N_2}{N_1+N_2-2\rho\sqrt{N_1N_2}}
\nonumber\\
& &\quad\qquad +\ts \frac{N_1^2+\rho^2N_1N_2-2\rho N_1\sqrt{N_1N_2}}
               {N_1+N_2-2\rho\sqrt{N_1N_2}}
\nonumber\\
&=&\beta\alpha P_1+N_1\,.
\eeqno
Hence, we have 
$$
h(Y|US)\geq \ts \frac{1}{2}
\log\left\{
    (2\pi{\rm e})\left(\beta{\alpha}P_1+ N_1\right)
    \right\}\,.
$$
Applying entropy power inequality to (\ref{eqn:zzooo}), 
we have
\beqno  
\ts \frac{1}{2\pi{\rm e}} 2^{2h(Z|US)}
&\geq& \ts \frac{1}{2\pi{\rm e}} 2^{2h(\tilde{Y}|US)}
      +\frac{1}{2\pi{\rm e}} 2^{2h(a{\tilde{\xi}_2})} 
\nonumber\\
&=& \beta\alpha P_1 + \tilde{N}_1 + a^2\tilde{N}_2
\nonumber\\
&=& \beta\alpha P_1+\ts \frac{(1-\rho^2)N_1N_2}{N_1+N_2-2\rho\sqrt{N_1N_2}}
\nonumber\\
&  & \quad\qquad  +\ts \frac{N_2^2+\rho^2N_1N_2-2\rho N_1\sqrt{N_1N_2}}
               {N_1+N_2-2\rho\sqrt{N_1N_2}}
\nonumber\\
&=&\beta\alpha P_1+N_2\,.
\eeqno
Hence, we have 
$$
h(Z|US)\geq \ts \frac{1}{2}
\log \left\{ 
(2\pi{\rm e})\left(\beta{\alpha}P_1+ N_2\right)
\right\}
\,.
$$
Thus the proof of Lemma \ref{lm:Rndau} is completed. 
\hfill\QED
%Finally we derive 

\end{document}